\documentclass[fleqn,usenatbib,useAMS]{mnras}

\usepackage{graphicx}
\usepackage[T1]{fontenc}
\usepackage[utf8]{inputenc}
\usepackage{ae,aecompl}          
\usepackage{newtxtext,newtxmath} 
\usepackage{xcolor}
\usepackage{tabularx}
\usepackage{tikz}  
\usepackage{comment}
\usetikzlibrary{positioning}
\usepackage{soul}

\graphicspath{{./plots/}}

\definecolor{dcrc}{rgb}{0.47,0.26, 0.41}

\newcommand{\dd}{\textrm{d}}

\title[Constraints on cosmologically coupled black holes]{Constraints on cosmologically coupled black holes from gravitational wave observations and minimal formation mass}

\author[L.~Amendola, D.~C.~Rodrigues, S.~Kumar, M.~Quartin]{
  Luca Amendola$^{1}$, 
  Davi C. Rodrigues$^{1, 2, 3}$,
  Sumit Kumar$^{4, 5}$,
  Miguel Quartin$^{3, 6, 7}$\\ 
  $^{1}$ Institut für Theoretische Physik, Universität Heidelberg, Philosophenweg 16, 69120. Heidelberg, Germany\\
  $^{2}$ Departamento de Física \& Cosmo-Ufes, Universidade Federal do Espírito Santo, 29075-910. Vitória\;-\;ES, Brazil\\
  $^{3}$ PPGCosmo, Universidade Federal do Espírito Santo, 29075-910, Vitória\;-\;ES, Brazil\\
  $^{4}$ Max-Planck-Institut für Gravitationsphysik (Albert-Einstein-Institut), D-30167 Hannover, Germany\\
  $^{5}$ Leibniz Universität Hannover, D-30167 Hannover, Germany\\
  $^{6}$ Instituto de Física, Universidade Federal do Rio de Janeiro, 21941-972, Rio de Janeiro, RJ, Brazil\\
  $^{7}$ Observatório do Valongo, Universidade Federal do Rio de Janeiro, 20080-090, Rio de Janeiro, RJ, Brazil
}

\date{\today}
\pubyear{2023}

\begin{document}
\label{firstpage}
\pagerange{\pageref{firstpage}--\pageref{lastpage}}
\maketitle

\begin{abstract}
    We test the possibility that the black holes (BHs) detected by LIGO-Virgo-KAGRA (LVK) may be cosmologically coupled and grow in mass proportionally to the cosmological scale factor to some power $k$, which may also act as the dark energy source if $k\approx 3$.
    This approach was proposed as an extension of Kerr BHs embedded in cosmological backgrounds and possibly without singularities or horizons. In our analysis, we develop and apply two methods to test these cosmologically coupled BHs (CCBHs) either with or without connection to dark energy.  We consider different scenarios for the time between the binary BH formation and its merger, and we find that the standard log-uniform distribution yields weaker constraints than the CCBH-corrected case. Assuming that the minimum mass of a BH with stellar progenitor is $2M_\odot$, we estimate the probability that at least one BH among the observed ones had an initial mass below this threshold.  
    We obtain these probabilities either directly from the observed data or by assuming the LVK power-law-plus-peak mass distribution.  In the latter case we 
    find, at $2\sigma$ level, that $k < 2.1$ for the standard log-uniform distribution, or $k < 1.1$ for the CCBH-corrected distribution. Slightly {weaker} bounds are obtained in the direct method. Considering the uncertainties on the nature of CCBHs, we also find that the required minimum CCBH mass value to eliminate the tensions for $k=3$ should be lower than 0.5 $M_\odot$ {(again at 2$\sigma$)}. {Finally, we show that future observations have the potential to decisively confirm these bounds.}
\end{abstract}


\section{Introduction} \label{sec:intro}

Recently, a new intriguing hypothesis about the origin of the cosmic acceleration has been put forward by \citet{Croker:2019mup, Croker:2020plg}, with further developments by \citet{Farrah:2023opk}. According to this scenario, black holes (BHs) grow in mass due to a form of cosmological coupling unrelated to local accretion. If this growth is fast enough, it could compensate the decrease in number density due to the cosmic expansion, and generate a form of effective cosmological constant. These BHs deviate from the standard Kerr solution.\footnote{Since astrophysical BHs are expected to have angular momentum, Kerr is a better description than Schwarzschild. Kerr BHs are  asymptotically Minkowski and have a singularity ``dressed'' by a horizon. Kerr-de Sitter solutions are also known \citep[for a review see][]{Akcay:2010vt}.} There is expectation that these solutions can be found within general relativity (GR) and that they could be singularity free \citep[see also][]{Dymnikova:2016nlb}. These non-standard BH solutions are asymptotically Friedmann-Roberton-Walker, rather than Minkowski \citep{Faraoni:2007es, Croker:2019kje, Croker:2019mup, Croker:2020plg}.  They could provide an average pressure that would constitute the entire amount of dark energy needed to explain the cosmic acceleration if the BHs have the necessary abundance. \citet{Farrah:2023opk} argue that this {may be} the case. In a companion paper, \citet{2023ApJ...943..133F} have found strong indication in favor of just such a cosmological growth of supermassive BHs in elliptical galaxies. This growth seems very difficult to explain in terms of the standard local growth channels of accretion.
{From different considerations, \citet{Gao:2023keg, Cadoni:2023lqe, Cadoni:2023lum} provide further support for cosmologically coupled black holes.}

This new BH solution is at the moment a conjecture, and in fact, criticisms on the above framework within GR have appeared.  
For instance, \citet{Avelino:2023rac} considers the use of gravastars as CCBHs, criticizes the mechanism for generating cosmological pressure assuming that the Birkhoff theorem can be applied and, additionally, points out that, if CCBH momentum is conserved, SMBHs could not be at rest with respect to their host galaxy. \citet{Parnovsky:2023wkc} criticizes both the uncertainty in the estimation of SMBH data and the possibility of BHs to provide a negative pressure. \citet{Mistele:2023fds} points out an inconsistency between the averaging process proposed by \citet{Croker:2019mup} and the action principle, thus this work also puts into question the proposed mechanism that generates the cosmological pressure. \citet{Wang:2023aqe} argue that a cosmological coupling cannot exist within general relativity
inside gravitationally bound systems.
Here we take an agnostic view and show how this may be at tension with GW data, independently of the detailed microphysics that may lead to the CCBH realization.

This paper is devoted to testing the cosmologically coupled BHs (CCBH) by looking at the current and future datasets from LIGO-Virgo-KAGRA (LVK). The current understanding is that the gravitational waves (GW) detected by LVK come from the merging of BHs with stellar progenitors. If the CCBH hypothesis is correct, they must have grown to the observed mass from an initially lower mass. However, BHs with a stellar progenitor cannot be formed with arbitrarily low masses. Observationally, there is evidence of a paucity of BH masses between $2-5 M_{\odot}$ \citep{2010ApJ...725.1918O, KAGRA:2021duu, deSa:2022qny}, while there is no conclusive evidence for a BH with mass about or below $2 M_\odot$ \citep[see also][]{LIGOScientific:2022hai}. 
Our main results are based on the conservative {mass threshold $m_{\rm th} =  2.0 M_\odot$} as the minimum BH formation mass. However, we also consider changes in different values for this threshold. {The higher (lower) is this threshold, the stronger (weaker) are our constraints. If in the future a stellar BH is detected with mass below $2 M_\odot$ this will clearly imply that $m_{\rm th}$ has to be smaller than what is detected, weaking our constraints. In particular, the constraints would vanish for very low thresholds of $\lesssim 0.5 M_\odot$.}
In Appendix \ref{app:minimumMass} we present further discussions on this.

In this paper, we 
use two complementary approaches, explore ways to alleviate the tensions\footnote{We use the word ``tension'' for rejections at a level higher than 2$\sigma$.} we find, and briefly discuss future prospects. We conclude that the CCBH as proposed by \citet{Farrah:2023opk} is in strong tension with what we know about stellar progenitor BHs, but there  still is an open parameter space where it can survive the present test. In particular, we find no relevant tension for the CCBH case studied by \citet{Croker:2021duf}. The forthcoming new GW datasets will soon shed further light on the CCBH conjecture.

The codes we used for this work are available at \url{https://github.com/itpamendola/CCBH-direct}  and \url{https://github.com/davi-rodrigues/CCBH-Numerics}.

\section{Cosmologically coupled black holes} \label{sec:review}

\citet{2023ApJ...943..133F} considered three samples of red-sequence elliptical galaxies at different redshifts and found that the growth of supermassive BHs is significantly larger than the growth of stellar mass, being a factor of 20 from $z\sim 2$ to $z\sim 0$. This growth is too large to be compatible with the expected  accretion rate \citep{2023ApJ...943..133F}. This suggests a different growth mechanism  such that  $m_{\rm BH} \propto m_* (1+z)^{-3.5 \pm 1.4}$, 
at $90\%$ confidence level, where $m_*$ is the stellar mass of the galaxy and $m_{\rm BH}$ the supermassive BH mass of the same galaxy. 

A possible explanation for the above physics comes from the proposal of cosmologically coupled BHs (CCBHs)~\citep{Faraoni:2007es, Croker:2019mup, Croker:2021duf}. In this case, BHs would grow following the parametrization \citep{Croker:2021duf} 
\begin{equation} \label{MkParametrization}
    m(a) = m(a_i) \left( \frac{a}{a_i}\right )^k \, ,
\end{equation} 
where $k\ge 0$ is a constant, $a_i$ is the cosmological scale factor at the time of the BH formation and $m(a_i)$ is its mass at that time.

\citet{Farrah:2023opk} explore the viability of  $k\approx 3$, which would both explain the supermassive BHs growth and provide a  source of dark energy capable of generating the observed $\Omega_\Lambda$ value. The latter requires further assumptions, in particular a proper star formation rate,  that all the remnants with mass $> 2.7$$M_\odot$ are BHs, and that all the BHs follow the above mass parametrization.  
Beside  the theoretical issues commented in Sec.~\ref{sec:intro}, \citet{Lei:2023mke} used JWST data and found a conflict with \citet{2023ApJ...943..133F} parametrization at redshifts $z \sim 4.5 - 7$. These high redshift results are mostly independent of our analysis: the constraints we find come from lower redshifts, as will be shown when constraining the maximum redshift of binary BH formation ($z_{\rm max}$). 

If all BHs are cosmologically coupled with $k=3$, \citet{Rodriguez:2023gaa} pointed out that this would be in contradiction with globular cluster NGC 3201 data, since it would imply that one of the BHs would have a mass below $2.2 M_\odot$. 
A similar test on two Gaia DR3 stellar-BH systems with reliable age estimation has been carried out by \citet{Andrae:2023wge}, resulting in the 2$\sigma$ upper limit $k\leq3.2$ assuming the same $2.2 M_\odot$ limit and fixing the background to $\Lambda$CDM. In Appendix \ref{app:minimumMass} we detail further aspects of the BH minimum mass in the context of CCBH.

\citet{Ghodla:2023iaz} found that the rate of mergers and their typical masses in a CCBH scenario would be hardly compatible with LVK observations; they also point out that CCBHs should exhibit  lower spins  due to their increase in mass. They have also derived a modified delay time for CCBHs that we consider in Sec.~\ref{sec:ghodla}.
This modified delay time makes CCBHs more incompatible with current data, as we develop here. 

Our purpose here is to test if the BHs detected from their coalescence waves could be cosmologically coupled. Before the results from \citet{Farrah:2023opk}, \citet{Croker:2021duf}  \citep[see also][]{Croker:2019kje} developed simulations of merging BHs and considered the impact of the cosmological coupling on the LVK detected BHs, showing that $k=0.5$ would be preferred over the standard $k=0$, at least for certain isolated-binary-evolution model. We use here the most recent data from LVK, together with more recent delay time expectations.  A crucial difference between this work and  the
works of \citet{Croker:2021duf, Ghodla:2023iaz} on CCBH and LVK data is that they started
from a given BBH formation mass, assumed to be realistic, and consider if they could mimic LVK data from that initial mass distribution. Here we aim to estimate what is the probability that at least one of the observed BHs via LVK would be {formed with a mass below a given threshold mass} \citep[thus in part similar to][]{Rodriguez:2023gaa}. A key quantity for modelling the CCBH effects on LVK data is the delay time $t_{\rm d}$ (i.e.,~the interval between BBH formation and merger), which is detailed in Sec.~\ref{sec:td}.

Within the general class of CCBHs, we distinguish two cases.
If BHs have dark energy implications and constitute its only source, as proposed in \citep{Croker:2020plg, Farrah:2023opk}, then the constant $k$, which parametrizes the energy density of BHs as a function of $a(t)$, has  a direct connection with the dark energy equation of state parameter $w$, with  
\begin{equation}
    k \equiv - 3 w \, .\label{kEquals3w}
\end{equation} 
We call this scenario dark energy BHs, DEBH.

If CCBHs masses increase following  Eq.~\eqref{MkParametrization}, but if they do not constitute a dark energy source, we call these growing BHs (GBHs). For instance, \citet{Croker:2021duf} considered BHs with this property with $k=0.5$. This picture can be realised if the CCBHs contribution to the total cosmological energy density is negligible, or if GBHs are receiving energy with another field, say an additional scalar field. In the GBH scenario, dark energy is fully sourced by a cosmological constant and there is no deviation from $\Lambda$CDM  background cosmology. 

The two models, DEBH and GBH, have identical background cosmological evolution  for $k=3$ but diverge otherwise. 
We consider both in the following.

\section{Delay time and the mass correcting factor} \label{sec:td}

The merging of binary black holes (BBH) systems detected by LVK is commonly considered to be the end of a pair of BHs that orbited together from several Myrs to several giga-years before the merger \citep{KAGRA:2021kbb}. 
These BHs masses are  consistent with them being remnants of star progenitors, and this constitutes the standard interpretation \citep[e.g.,][]{Belczynski:2016obo, Mapelli:2020vfa, vanSon:2021zpk, Chen:2022gaf, KAGRA:2021duu}.

The relevant time during which the CCBH effect \eqref{MkParametrization} is active extends between the BHs formation and their merger. We call this time the BBH delay time and denote it by $t_{\rm d}$. We note that another delay-time definition, as the time between the stellar pair formation and the BBH merger, is also used in the literature. However, they typically differ by a few Myr \citep{vanSon:2021zpk}, hence both definitions are essentially the same. 

For the GBH scenario, we consider any value of $k$ in the range $0\leq k \leq 3$ \citep{Croker:2021duf}, where $k=0$ corresponds to the standard case (uncoupled Kerr BHs).  Apart from the $k=3$ case, other values of particular interest that have been discussed in the literature  are $k=0.5$ \citep{Croker:2021duf} and $k=1$ \citep{Cadoni:2023lum}.

For the DEBH case, changing the $k$ value changes cosmology. For clarity, this case will be  parameterized with a constant $w$, instead of $k$. We mainly consider $ -0.6 \leq w \leq -1$. We do not consider more negative $w$ values since they can only strengthen our constraints. 
The DEBH scenario has no limit that leads simultaneously to standard BHs and standard background cosmology. 

\citet{KAGRA:2021kbb, KAGRA:2021duu} state that the distribution of delay times can be approximated by a log-uniform distribution (i.e., $p(t_d) \propto 1/t_d$) with $0.05 < t_d (\mbox{Gyrs}) < 13.5 $ for BBHs. It is also pointed out that the formation of the first BBHs is restricted to $z < 10$. This picture is in good agreement with simulations and observational constraints \citep[e.g.,][]{Belczynski:2016obo, vanSon:2021zpk, Fishbach:2021mhp}.  

CCBHs, on the other hand, grow with time and therefore their delay times will also change: since larger BH masses dissipate energy faster through GW emission, the delay time of CCBHs should be smaller than for ordinary BHs for the same initial mass and orbit \citep{Croker:2020plg, Farrah:2023opk,Ghodla:2023iaz}. This does not imply that the delay time distribution of the \textit{detected} BBH mergers will favour shorter times.
In particular, as commented by \citet{Farrah:2023opk, Ghodla:2023iaz}, BBHs that would not merge before $z=0$ in the standard picture may merge if CCBH is true.  We will explore in more detail the possible CCBHs changes to the $t_{\rm d}$ distribution in Sec.~\ref{sec:ghodla} and Appendix \ref{app:CCBHtdGeneral}. Of particular relevance, there it is shown that the CCBH corrections to the delay-time distribution favor larger delay times than the log-uniform distribution. Therefore, studying the log-uniform case is useful also because  it provides conservative bounds.

Due to such unknowns, we use a log-uniform distribution for $t_{\rm d}$ varying three parameters that have a direct impact on the $t_{\rm d}$ values ($t_{\rm min}, \, t_{\rm max}$ and $z_{\rm max}$, as detailed below). Moreover, in Appendix \ref{app:changing_td} we explore the possibility of a steeper PDF for the $t_{\rm d}$ distribution (i.e., smaller delay times on average), which reduces the tensions we find in the main analysis. 

We adopt then here the log-uniform $t_{\rm d}$ distribution, 
\begin{equation}
      \log t_{\rm d} \sim U(\log t_{\rm min}, \log t_{\rm x}), \label{tdDistribution}
\end{equation}
where $t_{\rm x }$ is the minimum between the maximum delay time $t_{\rm max}$ and the time difference between the merger redshift ($z_{\rm m}$)  and the  maximum redshift with BBH formation ($z_{\rm max}$). As \textit{reference values}, we consider
\begin{align} \label{referenceValues}
    &t_{\rm min}=0.05\, {\rm Gyr}, \nonumber \\
    &t_{\rm max}=13.0\, {\rm Gyr}, \nonumber \\
    &z_{\rm max}=10, \\
    & w=-1,\nonumber\\
    & k=3\, . \nonumber
\end{align}
Besides these reference values, we also explore other combinations.  We anticipate that the CCBH tension that we find here either increases or stays constant if $t_{\rm max}$, $z_{\rm max}$, or $t_{\rm min}$ are increased. For the cosmological model we assume $\Omega_{\rm m} = 0.32$ and $H_0 = 70$ km s$^{-1}$ Mpc$^{-1}$. 

\citet{Fishbach:2021mhp} studied a possible correlation between $t_{\rm d}$ and $m_1$ considering observational data. It was found a marginal preference for smaller masses values to have larger $t_{\rm d}$. We do not consider such a correlation here, but if future analyses confirm 
this mass delay-time correlation, it will result in stronger bounds on the CCBH model from GW data.
In any case, the true delay-time distribution and its dependence on the mass are still uncertain \citep[e.g.,][]{vanSon:2021zpk}. 

Let us now consider a BH that is formed at $z_{i}$ and merges after
a delay time $t_{d}$ at $z_{m}$. Then
\begin{equation}
    t_{d}=\int_{z_{i}}^{z_{m}}\frac{\dd z}{(1+z)H(z)} \, . \label{eq:td}
\end{equation}
This can be inverted  for a given $w$ using 
\begin{equation}
    H^{2}=H_{0}^{2}\big[\Omega_{m}(1+z)^{3}+(1-\Omega_{m})(1+z)^{3+3w}\big] \, ,
\end{equation}
leading to
\begin{equation}
    z_{i}=z_{i}(z_{m},t_{d}) \, .
\end{equation}
Then the initial mass $m_i$ of BHs will be a function of $z_{m},\, t_{d}, \, k$, and proportional to the mass $m_m$ at merging time,
\begin{equation} \label{MiMm}
    m_{i}=m_{m}\left(\frac{1+z_{m}}{1+z_{i}(z_{m},t_{d})}\right)^{k}.
\end{equation}

\bigskip

From a given set of observed BHs masses and a $t_{\rm d}$ distribution \eqref{tdDistribution}, we aim to find the probability that none of the observed BHs was formed with mass below {the mass threshold $m_{\rm th}$ (i.e., $M_i < m_{\rm th}$).} If this probability is close to 1, then there is no tension between observations and the CCBH model. Otherwise,  there is tension between the observational data, the model and the given assumptions. We will show that the latter is the case. Alternatively,  one should invoke significant changes in some of the basic assumptions, e.g.~a lower threshold for BH formation, shorter delay times, a late BH formation, or a different growth scheme. 

We estimate this probability in two complementary ways. In the first one, we use directly the current dataset and find the joint probability that at least one BH was born with a mass below threshold under the CCBH hypothesis. Since we do not correct the observed distribution for the selection effects, we implicitly discard low-mass BHs from the estimation, and therefore we end up with more conservative, but possibly more robust, estimates. We call this the {\it direct method}. 
In the second method, we derive the expected initial mass distribution of BHs taking into account the selection bias of the detectors through the power-law-plus-peak (PLPP) profile \citep{LIGOScientific:2020kqk, KAGRA:2021duu}. This method leads to stronger constraints. We denote this as the {\it PLPP method}.

\begin{figure}
	\begin{tikzpicture}
  		\node (img1)  {\includegraphics[width=0.45\textwidth]{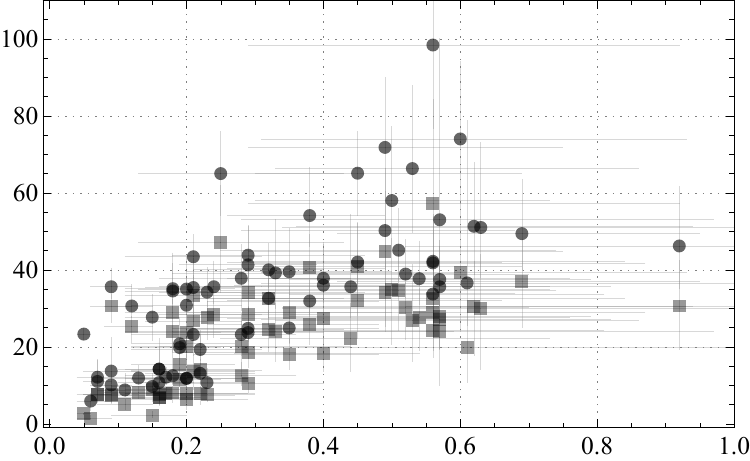}};
		\node[below=of img1, node distance=0cm, yshift=1.1cm, xshift=0.3cm, font=\color{black}] {\normalsize $z_m$};
		\node[left=of img1, node distance=0cm, rotate=90, yshift=-0.9cm, xshift=1cm] {\normalsize $m$ ($M_\odot$)};
	\end{tikzpicture} 
	\caption{The distribution of $m_1$ (disks) and $m_2$ (squares) as a function of the merging redshift ($z_m$). Data from Table \ref{tab:zXm}.}
	\label{fig:plotZxM.pdf} 
\end{figure}

The GWTC-3 data we use are shown in Table~\ref{tab:zXm} and in Fig.~\ref{fig:plotZxM.pdf}. These data come from confident BBH and NSBH events \citep{LIGOScientific:2021djp, KAGRA:2021duu} that satisfy $p_{\rm astro} > 0.5$ and FAR$_{\rm min}$ < 1 yr$^{-1}$.  In our analysis, we use separately either the primary $m_1$ masses or the secondary $m_2$ ones. Therefore, each selected mass corresponds to an independent history of a compact binary evolution and merger. Considering all the $m_1$ and $m_2$ in a single analysis would be incorrect since binary BHs  have the same $t_{\rm d}$ and would therefore not be independent. Although we consider here results with either primary or secondary masses, emphasis is given on the results for $m_1$ masses, since these produce more robust and more conservative constraints. For the direct method, this choice automatically removes BHs that are outliers with particularly low mass and have a large impact on the statistics used. For the PLPP method, the $m_1$ data is more robust since its distribution depends on one less parameter, with non-negligible uncertainty, than the $m_2$ distribution. In principle, one could consider $m_2$ masses for all the BBH cases, and change to $m_1$ masses for the NSBH systems, but the impact on the results is small since there are only one or two NSBH systems in our selected sample.

\begin{table}
    \centering
    \caption{
    Selected confident GW events from GWTC-3 catalog \citep{LIGOScientific:2021djp} that are classified in \citep{KAGRA:2021duu} as BBH or NSBH systems (we require $p_{\rm astro} > 0.5$ and FAR$_{\rm min}$ < 1~yr$^{-1}$), for a total of 72 events.  The columns show the event name, the merging redshift and the primary and secondary masses. The full table is provided electronically.} 
    \begin{tabularx}{0.48\textwidth}{l  X  r  X  r  X  r}
     \hline \hline
     Event & \hspace*{.3cm} & \multicolumn{1}{c}{$z_m$} & \hspace*{.3cm}& \multicolumn{1}{c}{$m_1 $ (M$_{\odot}$)} & \hspace*{.3cm} & \multicolumn{1}{c}{$m_2 $ (M$_{\odot}$)}\\
     \hline 
     \text{GW150914} & & $0.090_{-0.030}^{+0.030}$ & & $35.6_{-3.1}^{+5.}$ & & $30.6_{-4.}^{+3.0}$ \\[.1cm]
     \text{GW151012} & & $0.21_{-0.09}^{+0.09}$ & & $23._{-6.}^{+15.}$ & & $14._{-5.}^{+4.}$ \\[.1cm]
    \text{GW151226} & & $0.09_{-0.04}^{+0.04}$ & & $13.7_{-3.2}^{+9.}$ & & $7.7_{-2.5}^{+2.2}$ \\[.1cm]
    \text{GW170104} & & $0.20_{-0.08}^{+0.08}$ & & $31._{-6.}^{+7.}$ & & $20._{-5.}^{+5.}$ \\[.1cm]
    \text{GW170608} & & $0.070_{-0.020}^{+0.020}$ & & $11.0_{-1.7}^{+6.}$ & & $7.6_{-2.2}^{+1.4}$ \\
     ... & &... & &... & & ...\\
    \hline
    \end{tabularx}
    \label{tab:zXm}
\end{table}

The selected sample, Table~\ref{tab:zXm}, has only two systems classified as NSBH by LVK, namely: GW200105\_162426 and GW190917\_114630. This classification depends on the adopted minimum mass for BHs, and \citet{KAGRA:2021duu} consider $2.5 M_\odot$. When considering that the minimum mass is  $2M_\odot$, there remains a single  NSBH, GW200115\_042309. Excluding all events with secondary mass less than $5M_\odot$ as potential NS or outliers, we are left with 69 $m_2$ data points.

\section{Direct constraints from the observed events}  \label{sec:direct}

Here we discuss the direct method. The formation redshift that a BH of merging mass $m_{m}$ observed at $z_{m}$ should have to initially form with a given threshold mass $m_{\rm th}$ is given by Eq.~\eqref{MiMm} as
\begin{equation}
    z_{\rm th}=(1+z_{m})\left(\frac{m_{\rm m}}{m_{\rm th}}\right)^{1/k}-1 \,,
\end{equation}
and the corresponding delay time is 
\begin{equation}
    t_{\rm th}=t_{d}(z_{m},z_{\rm th})=t_{d}(z_{m},m_{m},m_{\rm th})\,.
\end{equation}
If the delay time is larger than $t_{\rm th}$, the BH would have formed with
a mass below the threshold. If $t_{\rm th}$ plus the merger age $t(z_m)$ is larger than the cut-off $t_{\rm max}$, we take $t_{\rm th}=t_{{\rm max}}-t(z_m)$. Analogously, if $z_{\rm th}$ is larger than, say, $z_{\rm max}=10$, we should cut it at $z_{\rm max}$ to prevent formation at an unrealistically early epoch.
Now, given a normalized delay-time distribution $\Psi(t_{d})$, the probability that a BH has formation mass above $m_{\rm th}$ is 
\begin{equation}
    p_{i}(z_{m},m_{m},m_{\rm th})=\int_{0}^{t_{\rm th}}\Psi(t_{d})\dd t_{d} \,.\label{eq:ptd}
\end{equation}
The combined probability of having $N$ BHs within the acceptable formation mass range $>m_{\rm th}$ is
\begin{equation}
    P(m_1>m_{\rm th})=\prod_{i}^{N}p_{i}
\label{eq:fullp}
\end{equation}
and therefore the probability of at least one below-threshold BH is $1-P$. In order to reject the CCBH hypothesis at given $k$, we should find a small $P$ for the currently observed BHs.  In other words, the $p$-value for rejecting the CCBH hypothesis is $p=P(m_1>m_{\rm th})$.

\begin{figure*}       
    \includegraphics[width=0.75\textwidth]{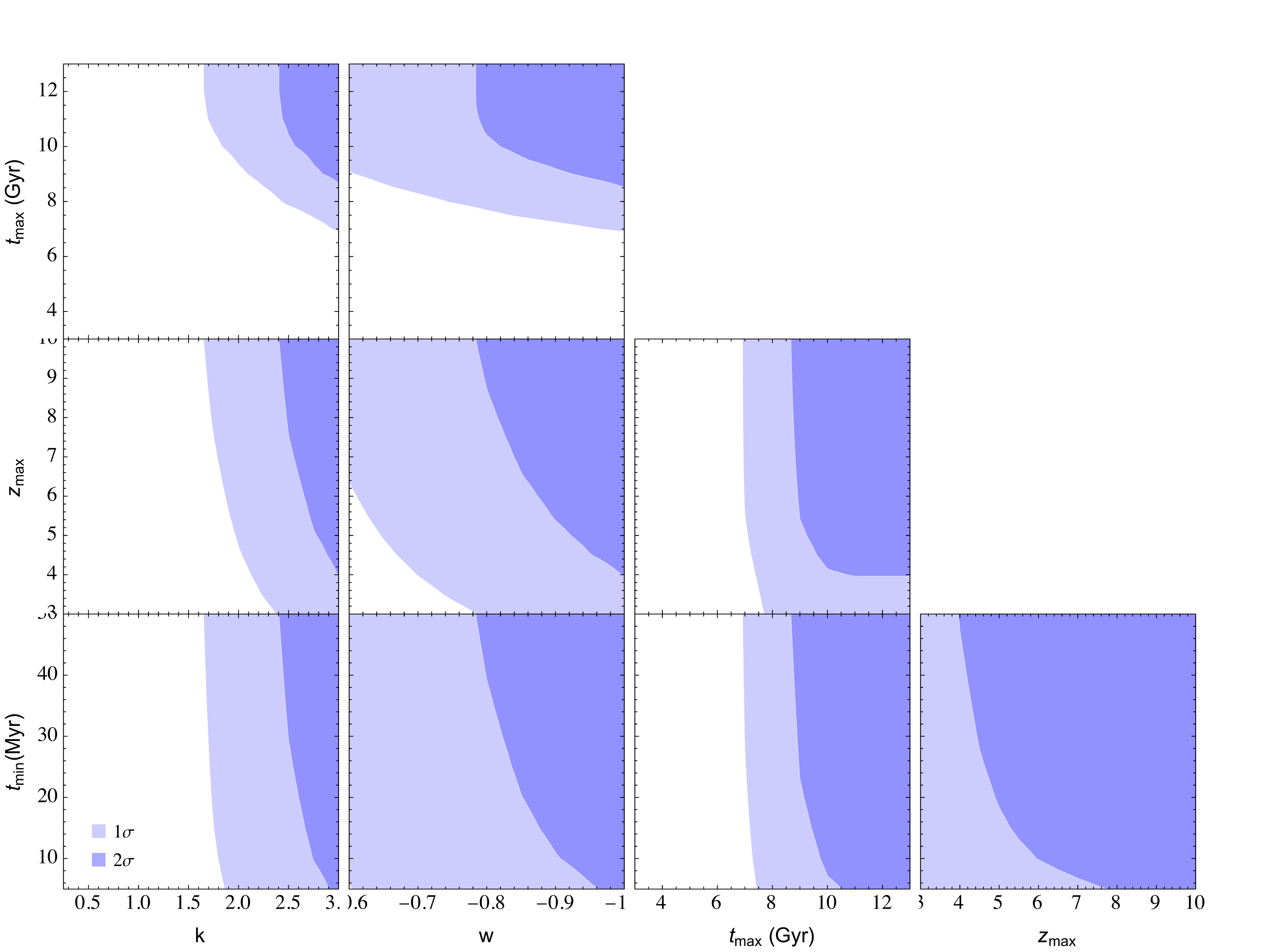}
    \caption{Direct method. Exclusion plots for the GBH and DEBH tensions.
    {Here we assume a minimum mass threshold  $m_{\rm th} = 2\, M_\odot$}. The reference values are always in the upper right corner of each plot. The first and second column correspond respectively to the GBH and the DEBH cases. The two last columns show results that are common to both approaches.}
    \label{fig:corner-plot-gbh} 
\end{figure*}

\begin{figure}
\begin{tikzpicture}
    \node (img1)  {\includegraphics[width=0.42\textwidth]{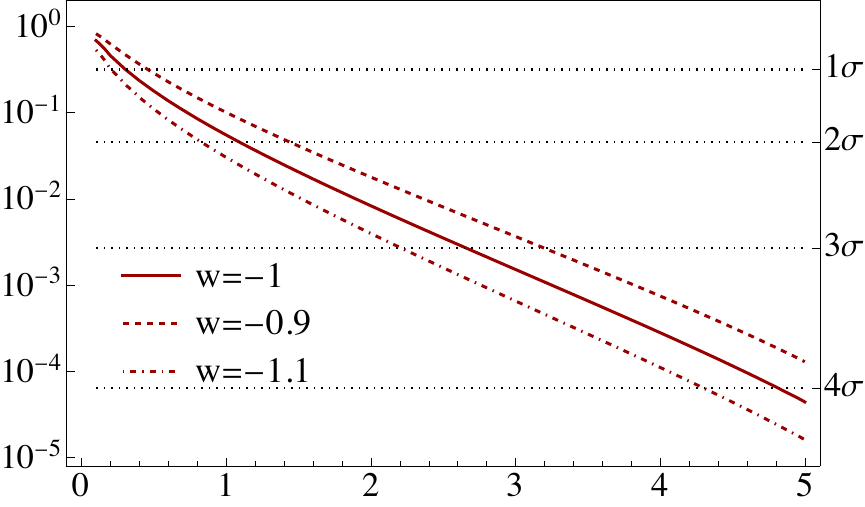}};
    \node[below=of img1, node distance=0cm, yshift=1.1cm, xshift=0.cm, font=\color{black}] {\normalsize Minimum mass threshold $m_{\rm th}$ ($M_\odot$)};
    \node[left=of img1, node distance=0cm, rotate=90, yshift=-.8cm, xshift=1cm] {\normalsize Probability };
    \end{tikzpicture}      
    \caption{
    Direct method.
    Plot of $P(m_1>  m_{\rm th})$  versus minimum BH mass in the DEBH scenario. The  dotted horizontal lines mark the $\sigma$ levels.} 
    \label{fig:min-mass-debh} 
\end{figure}

\begin{figure}	
    \begin{tikzpicture}
    \node (img1)  {\includegraphics[width=0.4\textwidth]{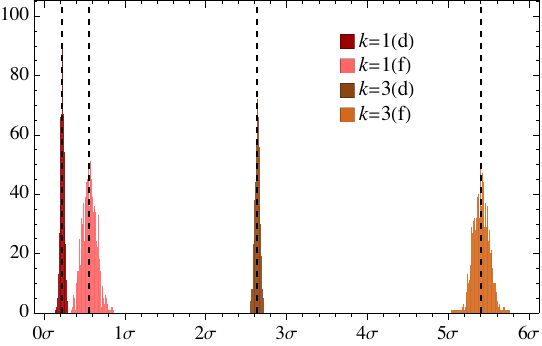}};
    \node[below=of img1, node distance=0cm, yshift=1.1cm, xshift=0.2cm, font=\color{black}] {\normalsize Tension};
    \node[left=of img1, node distance=0cm, rotate=90, yshift=-.7cm, xshift=0.7cm] {\normalsize Counts};
    \end{tikzpicture} 
    \caption{Direct method.  Tensions assuming {a fixed threshold $m_{\rm th} = 2 M_\odot$ and the reference $t_{\rm d}$ values~\eqref{referenceValues}}      
    for current data (d) and for the forecasted 250 LVK O4 events (f) obtained as  1000 random realizations of the current GW data, for $k=1$ and $k=3$. The vertical black dashed lines show the median value of each  distribution.} 
    \label{fig:simulation} 
\end{figure}

\begin{figure}
    \begin{tikzpicture}
    \node (img1)  {\includegraphics[width=0.43\textwidth]{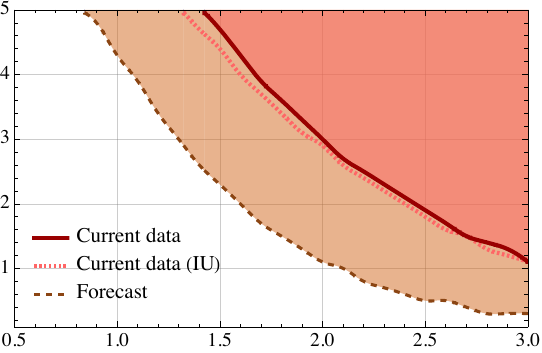}};
    \node[below=of img1, node distance=0cm, yshift=1.1cm, xshift=0.cm, font=\color{black}] {\normalsize $k$};
    \node[left=of img1, node distance=0cm, rotate=90, yshift=-.7cm, xshift=1.7cm] {\normalsize Minimum BH mass ($M_\odot$) };
    \end{tikzpicture} 
    \caption{
    Direct method. Excluded 2$\sigma$ regions  for current data (72 events, red) and forecast (250 events, brown). The GBH model is used here. The solid red line uses the observed mass and redshift distributions of all 72 events. The {dotted} red line is the resulting $2\sigma$ exclusion line based on the 72 events central values, {that is, ignoring uncertainties (IU)}. The forecast assumes no changes in the mass-redshift distribution and considers 250 events.
    } 
    \label{fig:plotkXmDirect} 
\end{figure}

Since GW observations pick preferentially high-mass BHs, the constraints we derive are on the conservative side.

We  consider both the DEBH case, in which the BH growth is linked to the dark energy so that $k=-3w$, and the alternative GBH scenario in which the BH growth does not influence the cosmological expansion. For simplicity, in this second case, we fix $w=-1$, i.e. the standard cosmological constant. In each case, there is then just one BH-cosmological parameter (in addition to the astrophysical ones, namely $t_{\rm min},\,t_{\rm max},\,z_{\rm max},\,m_{\rm th}$): either $k$ for the GBH case, or $w$ for the DEBH case.

 We also notice that the merger redshifts $z_m$ are obtained from the luminosity distance by assuming a $\Lambda$CDM evolution. However, in the DEBH scenario, the background is $\Lambda$CDM only for $w=-1$ so for any other value of $w$ we should derive a new set of $z_m$. This correction is however on average  $\Delta z=0.02$ for $w=-0.6$, and smaller for  $w$ closer to $-1$. This is negligible with respect to the current uncertainty in $z_m$, so we neglect it.

We illustrate in the corner plot Fig.~\ref{fig:corner-plot-gbh} the exclusion plots for various combinations of parameters for both scenarios,  implicitly fixing all the other parameters to the reference case {described above in Eq.~\eqref{referenceValues}.} In this and in the subsequent corner plot, the reference case (for which GBH and DEBH coincide) is always at the upper right corner; moving beyond this point  increases the rejection level of the CCBH hypothesis.

The main result is that, for DEBH and in the reference case, the probability of having no BHs below threshold is 0.0083, corresponding to 2.64$\sigma$. 
Using instead the $m_2$ masses, and excluding as potential outliers (perhaps neutron stars) the two compact objects with masses in the range  $2-5 M_\odot$, we obtain, as expected, a higher rejection level of 3.05$\sigma$. Decreasing $w$ into the phantom regime $w<-1$ makes the result stronger. For $w<-1.2$, using $m_1$ masses the rejection is at the 3$\sigma$ level (again, fixing all the other parameters to reference). For the other parameters, the range for which the tension is reduced below the 2$\sigma$ level are $t_{\rm max} < 8.7$ Gyr and $z_{\rm max} < 4$.

For $k \approx 3$, the dependence on the threshold mass is shown 
in Fig.~\ref{fig:min-mass-debh}. Using the references values, the tension is removed only if the minimum BH mass is lower than $1.1 M_\odot$ for $k =3 $. Within the DEBH scenario, that is, considering changes of the dark energy equation of state induced by $k = - 3 w$, the minimum BH mass should be below 1.4$M_\odot$ for $w = -0.9$, or below 0.9$M_\odot$ for $w = -1.1$. 

In Fig.~\ref{fig:simulation} we show the distribution of probabilities for 1000 realizations of the current data randomly chosen. 
The mass and redshift distribution for each event has been obtained from the posteriors samples of the latest GWTC data release \citep{LIGOScientific:2018mvr, LIGOScientific:2021usb, LIGOScientific:2021djp}. We use source frame mass distribution for each event.
 This narrow distribution shows that sticking with the best fit $z_m,\, m_m$ values is an acceptable approximation. For the forecast and $k=3$, the rejection level goes beyond 5$\sigma$. On the other hand, for $k=1$ even the forecasted data implies no relevant tension. 

In the same Fig.~\ref{fig:simulation} we  also estimate  $P(m_1>m_{\rm th})$ for a number of future events.  
After 4 months LVK O4 run has observed 44  significant BBH candidate events.\footnote{\url{https://gracedb.ligo.org/superevents/public/O4/}} If we further limit ourselves to those with preliminary FAR$_{\rm min} < 1 / {\rm yr}$ and preliminary BBH classification with over 90\% probability, 35 BBH candidates remain. Considering that O4 should last for 20 months, and that Virgo has not yet joined observations, one expects as a very conservative lower bound a total of 20/4 $\times$ 35 = 175 new BBH events (or 247 in total) by the end of O4. Therefore we  assumed 250 BBH events for our O4 forecasts in Fig.~\ref{fig:simulation}  and below. 

Fig.~\ref{fig:plotkXmDirect} explores the two most relevant parameters for this analysis, $k$ and minimum BH mass ($m_{\rm th}$), considering the observational distribution of the pair $(z_m, m_1)$ for each detected event. We use here the GBH picture, since the DEBH one changes significantly the cosmology if $k$ is not close to 3. The main result is the red solid line that delimits the $2\sigma$ excluded region using the current LVK data. To find this curve we proceed as follows: for each pair of values $(k, m_{\rm th})$ and a particular realization of the observational data distribution, we compute the probability that there is no BH with mass below $m_{\rm th}$. We repeat the previous evaluation for 300 realizations of the observational data distribution. A point in the $2\sigma$ curve corresponds to the 95\% quantile of the previous 300 realizations.

We stress the following results at $2\sigma$ level for the current and the forecast data: $i$) assuming $m_{\rm th} = 2 M_\odot$ for the minimum BH mass: $k < 2.5$ (current) and $k<1.6$ (forecast). $ii$) for $k = 3$, $m_{\rm th} < 1.1 M_\odot$ (current) and $m_{\rm th} < 0.3 M_\odot$ (forecast). For $k=1$, there are no constraints from the current data and the forecast yields a very weak $m_{\rm th}$ constraint. For $k = 0.5$, there are no constraints from this approach.

\begin{figure}
    \begin{tikzpicture}
    \node (img1)  {\includegraphics[width=0.42\textwidth]{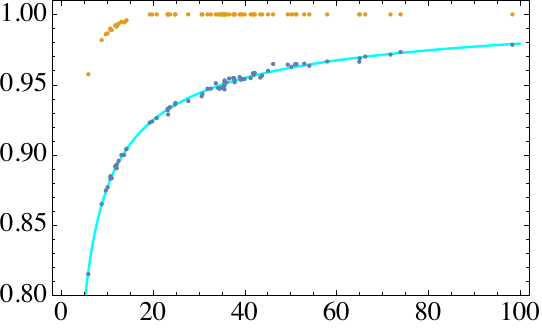}};
    \node[below=of img1, node distance=0cm, yshift=1.1cm, xshift=0.5cm, font=\color{black}] {\normalsize Mass ($M_\odot$)};
    \node[left=of img1, node distance=0cm, rotate=90, yshift=-.7cm, xshift=1.0cm] {\normalsize Probability};
    \end{tikzpicture} 
    \caption{
    Direct method. Individual probabilities for each BH to be formed with a mass above the {$2\,M_\odot$ threshold}, as a function of its observed mass at merging. {The cyan curve is the best fit given by Eq. (\ref{eq:bestfit}) for the reference case $k=3$.} The small dispersion about the curve is due to differences in the observed redshift.  The orange dots represent the probabilities for $k=1$; they are of course much closer to unity.}
    \label{fig:mass-ordered} 
\end{figure}

Finally, in Fig.~\ref{fig:mass-ordered} we plot the individual probabilities $p_i$ as a function of the BH mass for the reference case. As expected, small BHs are more likely to originate from below-threshold masses. However, all values of $p_i$ are relatively close to unity (larger than 0.8), implying that currently observed BH are more likely to be formed above the threshold than below it. It is the combined probability, rather that some peculiar outlier, that leads to the conclusion that at least one BH should have been formed below threshold. 

We find that, apart from a small dispersion that depends on the merging redshift ($z_m$), the probability $p(m)$ for a BH of observed mass $m$ (in solar mass units) can be very well approximated in the reference case by the following function
\begin{equation}\label{eq:bestfit}
    p(m) = C\log\big(A-Bm^{-1/2}\big)\,,
\end{equation}
with $A=26.72,B=30.88,C=0.3096$. The form of this function is suggested  by the analytical integration of Eq.~\eqref{eq:td} and Eq.~\eqref{eq:ptd} for a pure CDM model and a $1/t_d$ distribution; the coefficients are then obtained as a best fit to the actual $p_i$ values.

\section{Constraints using the power-law-plus-peak distribution} \label{sec:powerLawPlusPeak}

\begin{figure*}
  \centering
  \begin{tikzpicture}
    \node (img1) at (0,0) {\includegraphics[width=4.5cm]{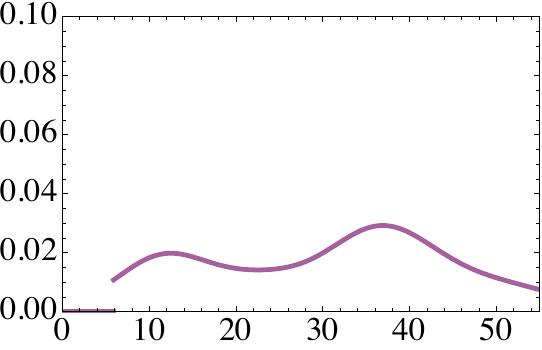}};
    \node[above=of img1, node distance=0cm, yshift=-0.6cm, xshift=0.0cm, font=\color{black}] {\bf Observed};
    \node[above=of img1, node distance=0cm, yshift=-1.1cm, xshift=0.0cm, font=\color{black}] {Influenced by detector bias};
    \node[below=of img1, node distance=0cm, yshift=1cm, xshift=0.2cm, font=\color{black}] {$m_1$ (M$_{\odot}$)};
    \node[left=of img1, rotate=90, node distance=0cm, yshift=-0.8cm, xshift=0.5cm, font=\color{black}] {PDF};

    \node (img2) at (6,0) {\includegraphics[width=4.5cm]{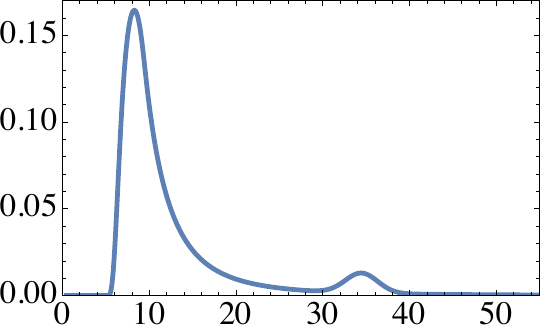}};
    \node[above=of img2, node distance=0cm, yshift=-0.6cm, xshift=0.0cm, font=\color{black}] {\bf Expected: merging};
    \node[above=of img2, node distance=0cm, yshift=-1.1cm, xshift=0.1cm, font=\color{black}] {Standard PLPP profile};
    \node[below=of img2, node distance=0cm, yshift=1cm, xshift=0.2cm, font=\color{black}] {$m_1$ (M$_{\odot}$)};
      
    \node (img3) at (12,2.4) {\includegraphics[width=4.5cm]{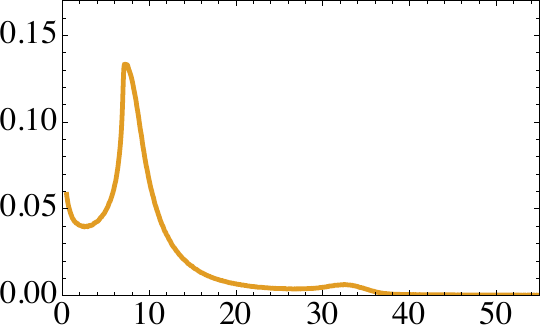}};
    \node[above=of img3, node distance=0cm, yshift=-0.6cm, xshift=0.0cm, font=\color{black}] {\bf Expected: formation ($k=3$)};
    \node[above=of img3, node distance=0cm, yshift=-1.1cm, xshift=0.0cm, font=\color{black}] {Modified PLPP profile};
    \node[below=of img3, node distance=0cm, yshift=1cm, xshift=0.2cm, font=\color{black}] {$m_1$ (M$_{\odot}$)};
    \node[left=of img3, rotate=90, node distance=0cm, yshift=-0.8cm, xshift=0.5cm, font=\color{black}] {PDF};
    
    \node (img4) at (12,-2.4) {\includegraphics[width=4.5cm]{plotArrowsMerger.pdf}};
    \node[above=of img4, node distance=0cm, yshift=-0.6cm, xshift=0.0cm, font=\color{black}] {\bf Expected: formation ($k=0$)};
    \node[above=of img4, node distance=0cm, yshift=-1.1cm, xshift=0.0cm, font=\color{black}] {Standard PLPP profile};
    \node[below=of img4, node distance=0cm, yshift=1cm, xshift=0.2cm, font=\color{black}] {$m_1$ (M$_{\odot}$)};
    \node[left=of img4, rotate=90, node distance=0cm, yshift=-0.8cm, xshift=0.5cm, font=\color{black}] {PDF};

    \draw[->, line width=0.5mm, darkgray] (img1) -- (img2);
    \draw[->, line width=0.5mm, darkgray] (img2) -- (img3);
    \draw[->, line width=0.5mm, darkgray] (img2) -- (img4);
  \end{tikzpicture}
  \caption{
    PLPP method. Illustration of the relation between three $m_1$ distribution contexts: the detected distribution, the expected distribution of all the $m_1$ masses from BBHs that merge (considering observational bias), and the expected $m_1$ distribution when it was formed. In the standard picture ($k=0, w=-1$), the distributions for formation and merging are the same. For CCBH, between formation and merger, BHs increase their mass, hence the formation distribution will favour lower masses than the standard picture. } 
    \label{fig:illustrationPPLP}
\end{figure*}

\subsection{General procedures}

We now move to the PLPP method. The true population of merged BBHs is not well described by the detected BBHs since detection bias has an important role. In particular, it is known that it is easier to detect massive BBH systems than low-mass systems: many low-mass BBH mergers are expected to happen but are undetected.

A successful profile for the mass distribution of merged BBHs (i.e., after modelling and correcting the detection bias) is the power-law-plus-peak (PLPP) one, as proposed by \citet{Talbot:2018cva} and analysed with current data \citep{LIGOScientific:2020kqk,KAGRA:2021duu}. Considering the $m_1$ mass distribution, the PLPP is a combination of a power law, described by $\beta(m_1)$, a Gaussian peak given by $G(m_1)$ and a smoothing function $S(m_1)$ that smooths the minimum mass probability transition.  The PLPP depends on seven parameters to describe the $m_1$ distribution: the power $\alpha$, the minimum and maximum masses ($m_{\rm min}, m_{\rm max}$), the Gaussian mean and standard deviation ($\mu, \sigma$), the smoothing parameter $\delta_m$ and the $\lambda$ parameter that adjusts the relative importance of the peak and the Gaussian. The peak is interpreted as a consequence of pair-instability supernovae \citep{Talbot:2018cva}. The smoothing function $S$ is introduced since the most probable $m_1$ values are not expected to be at the minimum $m_{\rm min}$: expectations from X-ray binaries and simulations \citep{Talbot:2018cva} suggest a smoother transition. Explicitly, the PDF reads,
\begin{align}
  &\pi(m_1)  \propto (1 - \lambda) \beta(m_1)S(m_1) + \lambda G(m_1)S(m_1)\, , \nonumber \\[.1cm]
  &\beta(m_1)  = \frac{\alpha -1}{m_{\rm min}^{1-\alpha} - m_{\rm max}^{1-\alpha}} m_1^{-\alpha}, \nonumber \\[.1cm]
  &G(m_1) = \frac{1}{\sqrt{2\pi} \, \sigma} \exp \left( - \frac{(m_1 - \mu)^2}{2 \sigma^2} \right) \, , \label{plppFunctions}\\[.1cm]
  &S(m_1) = 
    \begin{cases}
      \left[1 + \exp\left( \frac{\delta_m}{\delta m_1} - \frac{\delta_m}{\delta m_1-\delta_m} \right) \right ]^{-1}& \!\!\!\!\!\!, \, \delta m_1 < \delta_m\\[.5cm]
      1 &\!\!\!\!\!\!, \, \delta m_1 > \delta_m
    \end{cases}\, ,\nonumber
\end{align}
where $\delta m_1 \equiv m_1 - m_{\rm min}$. It is also imposed that $\pi(m_1) = 0$ for $m_1 < m_{\rm min}$ or $m_1 > m_{\rm max}$. {The PLPP parameter $m_{\rm min}$ states the minimum mass for both $m_1$ and $m_2$ masses. It should not be interpreted as stating the minimum mass of any BH, it is a fitted parameter that best describes the BBH merging population assuming that the PLPP profile holds. If this parameter is decreased, lower masses for the merging BBH population are possible; hence, in the CCBH context, even lower masses are necessary at formation time. Lowering $m_{\rm min}$ increases the constraints we find, while larger $m_{\rm min}$ values will decrease them. There is an analogous behaviour in the direct approach, in the sense that, the lower is the detected mass at merger time, the stronger will be our constraints.}

\begin{figure*}
    \begin{tikzpicture}
        \node (img1)  {\includegraphics[width=0.47\textwidth]{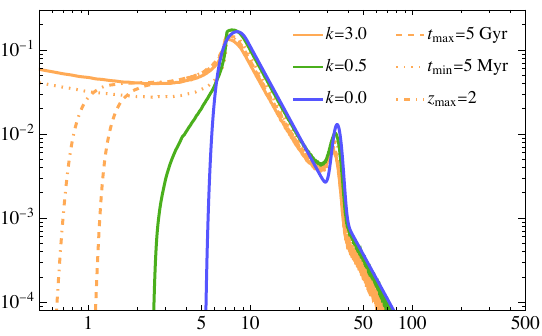}};
        \node[below=of img1, node distance=0cm, yshift=1.1cm, xshift=0.5cm, font=\color{black}] {\normalsize $m_1$ ($M_\odot$)};
        \node[left=of img1, node distance=0cm, rotate=90, yshift=-0.9cm, xshift=0.5cm] {\normalsize PDF};
    \end{tikzpicture} \hspace*{-.3cm}
    \begin{tikzpicture}
        \node (img1)  {\includegraphics[height=5.2cm, trim={0.6cm 0 0 0}, clip]{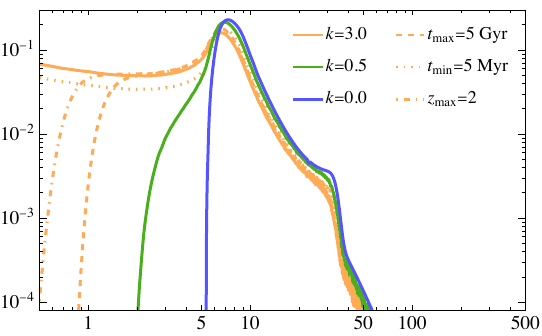}};
        \node[below=of img1, node distance=0cm, yshift=1.1cm, xshift=0.5cm, font=\color{black}] {\normalsize $m_2$ ($M_\odot$)};
    \end{tikzpicture} 
    \caption{PLPP method. The $m_1$ (left plot) and $m_2$ (right plot) distributions at formation for different parameter values, assuming the PLPP distribution at merger (see  
    Fig.~\ref{fig:illustrationPPLP}, using PLPP parameters of Eq.~\eqref{plppParametersBest}). In blue we show the $k=0$ case (no cosmological coupling). The green line shows the $k=0.5$ case, within the GBH approach. The solid orange curve  show the case $k=3$ with the reference values Eq.~\eqref{referenceValues}; while the other three orange curves show variations with respect to the latter values: $t_{\rm max} = 5$ Gyr (dashed), $t_{\rm min} = 5$ Myr (dotted) and $z_{\rm max} = 2$ (dot-dashed). The plots were generated with $10^7$ simulated events.}
    \label{fig:histFormationDist} 
\end{figure*}
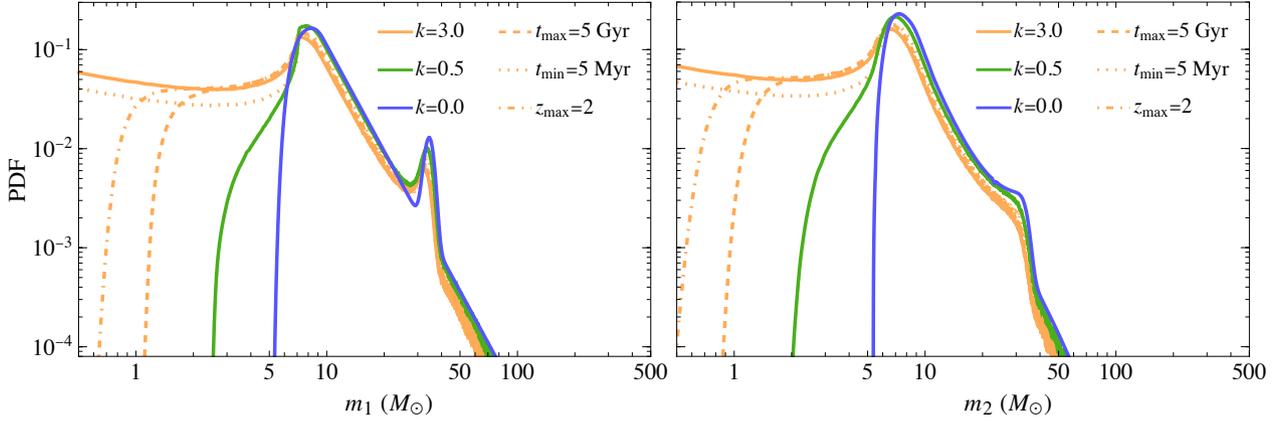

The parameters are found from GW observational data and considering the detector bias, through a hierarchical Bayesian approach \citep{KAGRA:2021duu}. The PLPP model represents the source frame mass distribution corrected for the selection effects. 
For the GWTC-3 data, the eight parameters that describe the $m_1$ and $m_2$ distributions are  \citep{ligo_scientific_collaboration_and_virgo_2023_7843926} (90\% credible intervals): 
\begin{alignat}{3}
  & \alpha = 3.40_{-0.49}^{+0.58},  &&\delta_m = 4.8_{-3.2}^{+3.3}\, {\rm M}_\odot, \nonumber \\
  & m_{\rm min} = 5.08_{-1.5}^{+0.87} \, {\rm M}_\odot,   \; \; && m_{\rm max} = 86.9_{-9.4}^{+11.} \, {\rm M}_\odot,  \label{plppParameters}\\
  & \mu = 33.7_ {-3.8}^{+2.3} \, {\rm M}_\odot, && \sigma = 3.6_{-2.1}^{+4.6} \, {\rm M}_\odot \, ,\nonumber\\
  &\lambda = 0.039_{-0.026}^{+0.058} \, ,  &&  \beta_q = 1.1_{-1.3}^{+1.8} \, .\nonumber
\end{alignat}
The $\beta_q$ parameter,  only needed to describe the $m_2$ distribution, will be discussed later on. 
The central values above are the medians of the posteriors, while  the uncertainties represent the $5\%$ and $90\%$ quantiles of the posteriors. From \citep{ligo_scientific_collaboration_and_virgo_2023_7843926} we also infer the maximum likelihood values, which read,
\begin{align}
    &\alpha = 3.55, \; m_{\rm min} = 4.82 {\rm M}_\odot, \; m_{\rm max} = 83.14 {\rm M}_\odot \, , \nonumber \\
    &\delta_m = 5.45 {\rm M}_\odot, \; \mu = 34.47 {\rm M}_\odot, \; \sigma = 1.87{\rm M}_\odot \, , \label{plppParametersBest}\\
    &\lambda = 0.019, \; \beta_q = 0.76 \, .\nonumber
\end{align}

\begin{figure}
    \begin{tikzpicture}
        \node (img1)  {\includegraphics[width=0.45\textwidth]{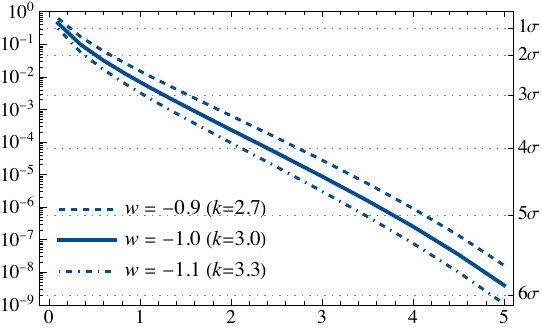}};
        \node[below=of img1, node distance=0cm, yshift=1.1cm, xshift=0.3cm, font=\color{black}] {Minimum mass threshold $m_{\rm th}$ ($M_\odot$)};
        \node[left=of img1, node distance=0cm, rotate=90, yshift=-0.9cm, xshift=0.9cm] {Probability};
    \end{tikzpicture} 
    \caption{PLPP method.  
   Probability that the DEBH model is in agreement with the minimum BH mass threshold, as a function of the latter, and for different $w$ values. This case uses $k = -3 w$.}	\label{fig:plotProbabilityNull} 
\end{figure}

To test the CCBH hypothesis, we use the merging BBH distribution, as provided by PLPP, to find the expected BBH mass distribution by the formation time, as illustrated in Fig.~\ref{fig:illustrationPPLP}. More precisely, let $M_{1, m}$ be a random realization of the PLPP distribution, where the index $m$ stands for merger time, and let $F_{z_m}$ be a random realization of the mass factor correction of eq.~\eqref{MiMm} at given $z_m$. Then the $m_1$ distribution at BBH formation time is the distribution of the random variate $M_{1, i}$, with
\begin{equation} \label{MFM}
    M_{1,i} = F_{z_m} M_{1, m} \, .
\end{equation}
Our results are found using at least $10^5$ realizations of each random variable. 

A \textit{caveat} of the above procedure is that the mass factor distribution depends on $z_m$, hence the $m_1$ distribution by formation time is $z_m$ dependent. This dependence on $z_m$ can be either considered by using the $z_m$ values from observations or by using a mean $z_m$ value as an approximation. The dependence on $z_m$ is weak (see also Fig.~\ref{fig:mass-ordered}), thus simply using an average $z_m$ value is sufficient. The quality of the approximation can be directly verified by computing any probability at the minimum and maximum $z_m$ values.

From the $m_1$ mass distribution at the BBH formation, one can compute what is the probability that one of the merged BBHs could have been formed with $m_1$ larger than a given mass threshold $m_{\rm th}$, denoted by $p(m_{\rm th}, z_m)$. For the various scenarios studied here, this probability for a single event with a mass larger than $2 M_\odot$ is not far from, but clearly below,  unity. Since the total number of merged BBHs is larger than the total number of confidently detected mergers, a minimum bound can be found by using the confidently detected BHs from gravitational waves (denoted by $N$), which we will use here. Hence, the probability of a given CCBH realization being compatible with existing data, similarly to eq.~\eqref{eq:fullp}, is  
\begin{equation} \label{pm1PLPP}
    P (m_1 > m_{\rm th}) = \prod_{j}^N p_j (m_{\rm th}, z_{m,j}) \approx p^N (m_{\rm th}, \overline{z_m}) \, ,
\end{equation}
where $\overline{z_m}$ is an average over all the $z_{m,j}$ values. The above equation can be computed either using the  redshift values of the observed merged BBH or by using the last approximation, which only depends on $\overline{z_m}$. The numerical differences are small, being  negligible when stating the tension with $\sigma$ units. 

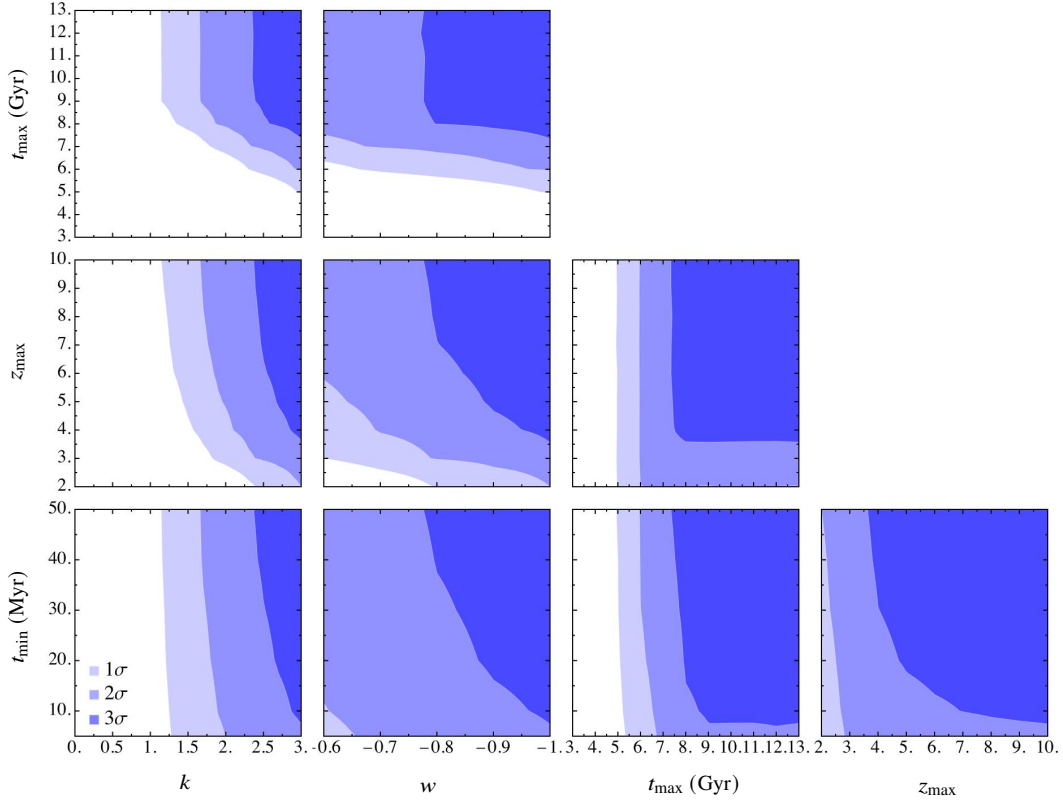
\begin{figure*}
  \begin{tikzpicture}
    \node (plot1)  {\includegraphics[width=0.75\textwidth]{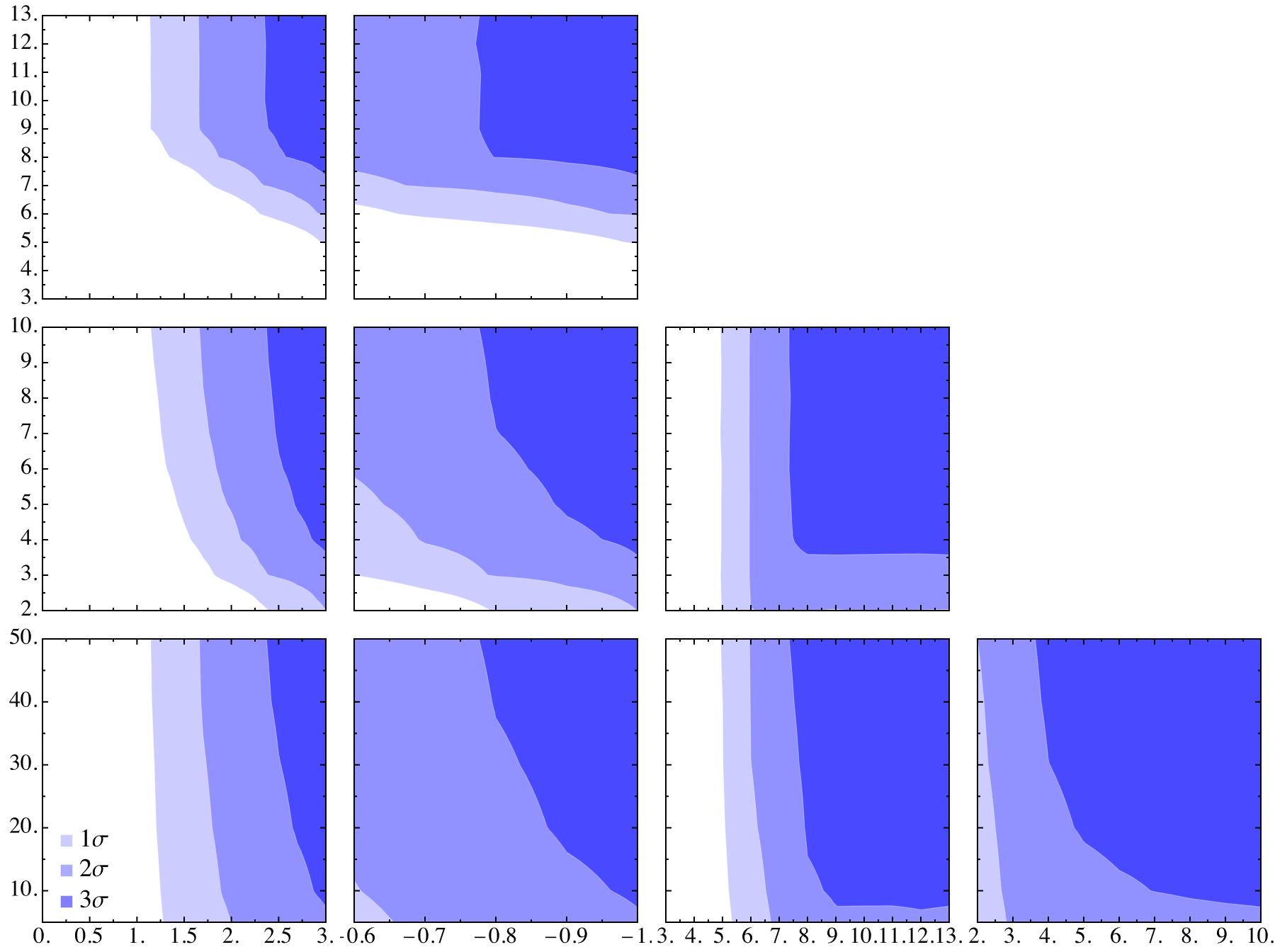}};
    \node[left=0.2cm of plot1, yshift=4.2cm, rotate=90]{ $t_{\rm max} \, {\rm (Gyr)}$};
    \node[left=0.2cm of plot1, yshift=0.6cm, rotate=90]{ $z_{\rm max}$};
    \node[left=0.2cm of plot1, yshift=-2.4cm, rotate=90]{ $t_{\rm min} \, {\rm (Myr)}$};
    \node[below=0.0cm of plot1, xshift=-4.8cm]{ $k$};
    \node[below=0.0cm of plot1, xshift=-1.6cm, yshift=-0.1cm]{ $w$};
    \node[below=0.0cm of plot1, xshift=1.9cm]{ $t_{\rm max} \, {\rm (Gyr)}$};
    \node[below=0.0cm of plot1, xshift=5.1cm, yshift=-0.1cm]{ $z_{\rm max}$};
  \end{tikzpicture} 
  \caption{PLPP method. {Same as Fig.~\ref{fig:corner-plot-gbh}.} The GBH and DEBH tensions with observational data, assuming that the PLPP distribution models the detection bias {and $m_{\rm th} = 2.0 M_\odot$}. The reference values are always in the upper right corner of each plot. The first and second columns correspond respectively to the GBH and the DEBH cases. The two last columns show results that are common to both approaches.}
  \label{fig:cornerPlotCCBHplpp} 
\end{figure*}

\begin{figure}
	\begin{tikzpicture}
  		\node (img1)  {\includegraphics[width=0.45\textwidth]{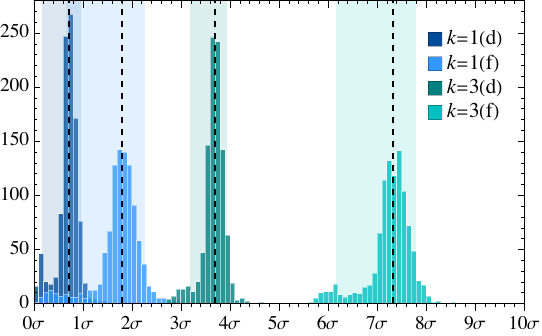}};
		\node[below=of img1, node distance=0cm, yshift=1.1cm, xshift=0.2cm, font=\color{black}] {\normalsize Tension};
		\node[left=of img1, node distance=0cm, rotate=90, yshift=-.9cm, xshift=0.70cm] {\normalsize Counts};
	\end{tikzpicture} 
    \caption{
    Similar to Fig.~\ref{fig:simulation} for the PLPP method. The histograms show the CCBHs tensions computed from $10^3$ different PLPP parameter realizations, including correlations \eqref{plppParameters}.   
    The median agrees with the reference values used for the PLPP parameters \eqref{plppParametersBest}. The rectangular light-colored regions delimit the 5\% and 95\% quantiles of the corresponding distributions, thus propagating the uncertainty in the PLPP parameters.} 
     
    \label{fig:histRandomPLPP}
\end{figure}

\begin{figure}
	\begin{tikzpicture}
  		\node (img1)  {\includegraphics[width=0.44\textwidth]{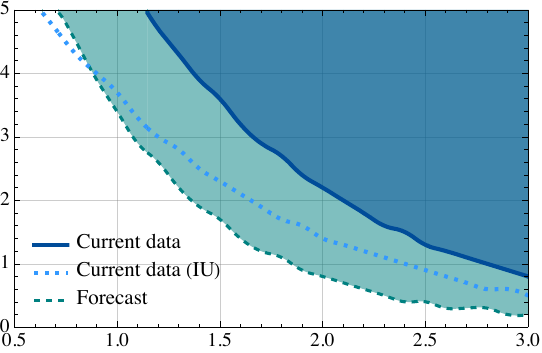}};
		\node[below=of img1, node distance=0cm, yshift=1.1cm, xshift=0.0cm, font=\color{black}] {\normalsize k};
		\node[left=of img1, node distance=0cm, rotate=90, yshift=-.7cm, xshift=1.90cm] {\normalsize Minimal BH mass ($M_\odot$)};
	\end{tikzpicture} 
    \caption{  
    {Similar to Fig.~\ref{fig:plotkXmDirect} for the PLPP method. Excluded regions at 2$\sigma$ for either the current data (72 events, blueish and solid) or for the forecast (250 events, greenish and dashed), using the GBH approach. The solid and dashed lines consider the full distribution of the PLPP parameters, while the dotted line ignores the uncertainties (IU) and use the best-fit PLPP parameters.}
    } 
    \label{fig:plotkXmPLPP}
\end{figure}

The above analysis applies for $m_1$ masses. Following \citet{Talbot:2018cva}, the $m_2$ distribution is given by the following conditional probability,
\begin{equation}
    \pi(m_2|m_1) \propto \left ( \frac{m_2}{m_1}\right )^{\beta_q} S(m_2) \Theta(m1-m_2)\, ,
\end{equation}
where $\beta_q$ is the single PLPP parameter that only appears in the $m_2$ distribution, $S$ is the smoothing function as defined in \eqref{plppFunctions} and $\Theta$ is the Heaviside theta function. In order to find the PDF $\pi(m_2)$ we marginalize over $m_1$,
\begin{equation} \label{plppM2pi}
    \pi(m_2) = \int_{m_{\rm min}}^{m_{\rm max}} \pi(m_2|m_1) \pi(m_1) \dd m_1 \, , 
\end{equation}
where $\pi(m_1)$ is given in eq.~\eqref{plppFunctions} and $\pi(m_2|m_1)$ needs to be normalized for each $m_1$ value. With this result, one can find the initial mass distribution and compute the probability $P(m_1 > m_{\rm th})$, eq.~\eqref{pm1PLPP}, for $m_2$ masses.

\subsection{Results} 

Using the approach illustrated in Fig.~\ref{fig:illustrationPPLP}, in Fig.~\ref{fig:histFormationDist} we show $m_1$ and $m_2$ distributions at formation time. Different values of $k$ and of the three parameters related to the $t_{\rm d}$ distribution ($t_{\rm max}$, $t_{\rm min}$, $z_{\rm max}$) are considered. Changing the latter three parameters can reduce the probability of a BH formation with mass below $2M_{\odot}$, but, as long as this PDF is not negligible for $m < 2M_{\odot}$, the tension between observational data and the minimum BH mass will increase with the number of detected BHs.  This figure also shows that the $k=0.5$ case, which was specially studied by \citet{Croker:2021duf} as a GBH model, is the safest among the shown cases in this figure.

By applying the mass factor correction on the PLPP distribution \eqref{MFM}, with the best-fit parameters \eqref{plppParametersBest}, and using eq.~\eqref{pm1PLPP} with $N=72$, the probability that no merged BBH was formed with mass smaller than $2 M_\odot$ is $P \approx 2 \times 10^{-4}$, thus implying a minimum tension of 3.7$\sigma$ for the reference values. 
With new detections in the current LVK run (O4), assuming that the PLPP profile with the best parameter values and the $t_{\rm d}$ distribution will remain the same, the tension is forecast to increase beyond $5\sigma$ (250 events). 

Instead of $m_1$, one may consider the $m_2$ masses, which lead to stronger constraints but depend on an additional PLPP parameter, $\beta_q$. In our events selection, Table~\ref{tab:zXm}, there are 69 $m_2$ masses larger than $5 M_\odot$: these are too massive to be NSs.  Using the $m_2$ distribution, eq.~\eqref{plppM2pi}, we find that the tension becomes 4.0$\sigma$ for $m_{\rm th} = 2 M_\odot$. Fixing $k=3$ but reducing the minimum mass, one finds that the tension disappears (i.e., less than 2$\sigma$) if $m_{\rm th} < 0.5 M_\odot$. These results assume the best-fit PLPP parameters. In the following, we return to focus on the $m_1$ analysis. 

In Fig.~\ref{fig:plotProbabilityNull} we show how the tension changes with the 
BH minimum mass, considering $m_1$ data. This figure also considers changes in $w$, in the context of the DEBH model. The tension decreases for larger $w$ values, which implies lower $k$ values. It is important to point out that the case $k=0.5$ is safe \citep{Croker:2021duf}: the forecast result implies no tension, as expected, since the modified PLPP profile at formation time, Fig. \ref{fig:histFormationDist}, show no relevant probability for $m_1 < 2 M_\odot$.

In Fig.~\ref{fig:cornerPlotCCBHplpp} we show exclusion plots for the DEBH and GBH models considering parameter variations with respect to the reference values (for the 72 observed $m_1$ data), fixed PLPP parameters as the best values and $m_{\th} = 2 M_\odot$. All the reference value changes here considered are such that the tension  decreases. This figure should be compared with Fig.~\ref{fig:corner-plot-gbh}. Both figures show qualitatively the same behaviour, but quantitatively, as expected, the PLPP approach is stronger.  
{We are thus left with only four parameters: $k$, $t_{\rm min}$, $t_{\rm max}$ and $z_{\rm max}$. We see that, out of these four, the two most important ones which could be changed to alleviate the tension are $k$ (as expected) and $t_{\rm max}$. However,} {we do not have a physical explanation for using a low $t_{\rm max}$ value (that would require an unexpected  restriction on the possible initial conditions of BBHs).}
For the DEBH model, the tension is reduced by increasing $w$, which corresponds to decreasing $k$, but even the case $w=-0.6$, which is far from the standard cosmological model, is not sufficient to drop the tension to an acceptable level. In the following, using Fig.~\ref{fig:cornerPlotCCBHplpp}, we highlight the necessary individual parameter ranges that reduce the reference tension from 3.7$\sigma$ to an acceptable level, below 2.0$\sigma$. They are: $k \leq 1.7$, $t_{\rm max} < 6$ Gyr, $z_{\rm max} < 2$. Reducing $t_{\rm min}$ from 50 Myr to 5 Myr only marginally improves the picture. Although imposing a strong bound on the maximum delay time value is effective at reducing the tension, there is no clear physical mechanism for this. In principle, BBHs are formed with a large spread on the starting orbit and masses. {In the standard picture ($k=0$), most of them need not to merge within a Hubble time \cite[e.g.,][]{Ghodla:2023iaz}.}

In Fig.~\ref{fig:histRandomPLPP} we consider how the $m_1$-based CCBH tension changes by considering $10^3$ samples of the PLPP parameter distribution \citep{ligo_scientific_collaboration_and_virgo_2023_7843926} and $m_{\th} = 2 M_\odot$. We stress the following properties: $i$) For the $k=1$ case, the results for the current and forecast data are compatible with no tension (i.e., below 2$\sigma$); $ii$) for $k=3$ case with current data, 95\% of the realizations have tension larger than 3.2$\sigma$; $iii$) for the forecast with $k=3$, 95\% of the realizations have tension larger than $6 \sigma$. $iv)$  Although the spread for the PLPP method is larger than for the direct one, the PLPP results are clearly stronger for $k=3$.

Figure~\ref{fig:plotkXmPLPP}, similarly to Fig.~\ref{fig:plotkXmDirect}, explores the combined variation of the two most relevant parameters for this analysis, $k$ and minimum BH mass ($m_{\rm th}$), and takes into account the full distribution of the PLPP parameters (instead of only using the best-value parameters).  The main result is the blue solid line that delimits the $2\sigma$ excluded region for the current data. To find this curve we proceed as follows: for each pair of values $(k, m_{\rm th})$ and a particular realization of the PLPP parameters distribution, we compute the probability that there is no BH with mass below $m_{\rm th}$. We repeat the previous evaluation for 300 realizations of the observational data distribution. A point in the $2\sigma$ curve corresponds to the 95\% quantile of the previous 300 realizations.

We stress the following results at $2\sigma$ level for the current and the forecast data: $i$) assuming $m_{\rm th} = 2 M_\odot$ for the minimum BH mass: $k < 2.1$ (current) and $k<1.4$ (forecast). $ii$) for $k = 3$, $m_{\rm th} < 0.8 M_\odot$ (current) and $m_{\rm th} < 0.2 M_\odot$ (forecast). For $k=1$, there are no constraints from the current data and the forecast yields $m_{\rm th} < 3.4 M_\odot$. For $k = 0.5$, there are no constraints.


\section{Modified delay-time distribution from CCBH physics}\label{sec:ghodla}

\begin{figure*}
	\begin{tikzpicture}
  		\node (img1)  {\includegraphics[width=0.45\textwidth]{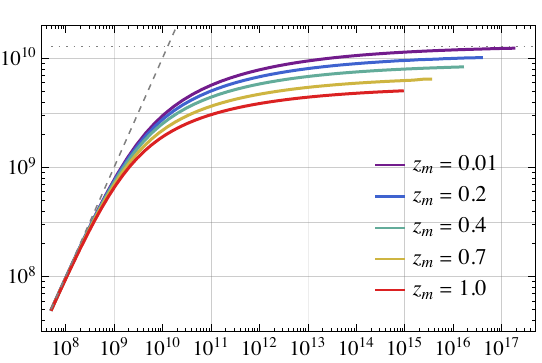}};
		\node[below=of img1, node distance=0cm, yshift=1.1cm, xshift=0.3cm, font=\color{black}] {$\tilde t_{\rm d}$ (years)};
		\node[left=of img1, node distance=0cm, rotate=90, yshift=-0.9cm, xshift=0.9cm] {$t_{\rm d}$ (years)};
	\end{tikzpicture} 
	\begin{tikzpicture}
  		\node (img2)  {\includegraphics[width=0.42\textwidth]{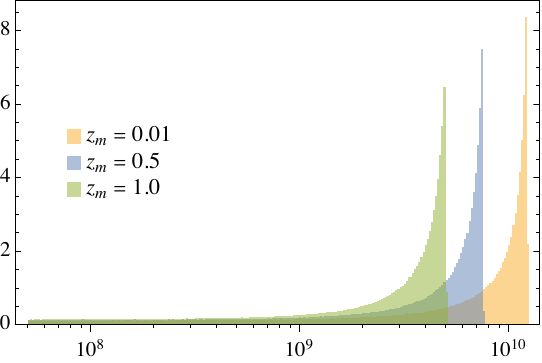}};
		\node[below=of img2, node distance=0cm, yshift=1.1cm, xshift=0.3cm, font=\color{black}] {$t_{\rm d}$ (years)};
		\node[left=of img2, node distance=0cm, rotate=90, yshift=-0.8cm, xshift=0.6cm] {PDF};
	\end{tikzpicture}
	\caption{\textit{Left.} The relation between the physical delay time $t_{\rm d}$ within the CCBH picture, with $k=3$, and an auxiliary one, $\tilde t_{\rm d}$, for $k=0$, at different merger redshifts $z_m$ (see eq.~\eqref{GhodlaRelationzm}). The curves stop at their maximum possible $t_{\rm d}$ value, such that all BBHs are formed at $z<10$. The horizontal dotted line is the maximum $t_{\rm d}$ value for $z_m = 0.01$. The dashed straight line satisfies $t_{\rm d} = \tilde t_{\rm d}$. \textit{Right.} Using the log-uniform distribution for $\tilde t_{\rm d}$, the $t_{\rm d}$ distribution is found as shown in this plot for different $z_m$ values. The modification in the delay time distribution, eq.~\eqref{GhodlaRelationzm}, strongly favours larger $t_{\rm d}$ values.}
	\label{fig:ghodla} 
\end{figure*}
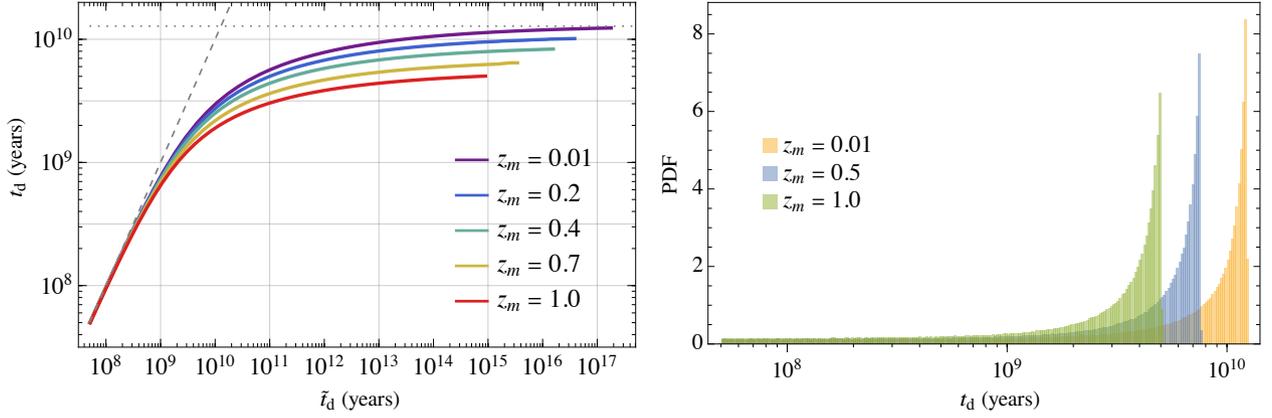

In a recent work, \citet{Ghodla:2023iaz} (henceforth G23) considered the mass increase of CCBHs in BBHs systems. As discussed in  Appendix \ref{app:CCBHtdGeneral}, even if individual delay times are significantly different due to the CCBH correction, this does not necessarily imply a change of the $t_{\rm d}$ distribution. Here we evaluate in detail the consequences within the picture described by G23, showing that this picture indeed leads to changes in the $t_{\rm d}$ distribution, but such changes can only strengthen the constraints we found from the log-uniform case. In the Appendix \ref{app:CCBHtdGeneral}, the reason for this behaviour is further explored. G23 find an explicit relation between the delay time of CCBHs with $k>0$, denoted as $t_{\rm d}$, and an auxiliary delay time that corresponds to the case without cosmological coupling ($k=0$), here denoted by $\tilde t_{\rm d}$. As  pointed out by  \citet{Croker:2019mup, Croker:2021duf}, there are two  factors that are important for the BBH dynamics and that are not part of the standard BH picture ($k=0$), namely: $i$) the emission of GWs becomes stronger with time,  since the system is increasing its mass and, $ii$) conservation of orbital angular momentum. For a given BBH with a given initial orbit at formation, these two factors make $t_{\rm d}$ smaller than the corresponding $\tilde t_{\rm d}$. The point $i$ above is quite clear and intuitive (since more massive binary systems should emit more GWs), the second point is not as obvious as it may seem, and we briefly review this subject. 

The issue with angular momentum conservation can be traced to the question: as a CCBH increases its mass proportionally to $a^k$, should its velocity continue the same as if it had no cosmological mass increase? Unless further details in the microphysics are specified, one cannot present a definitive answer, because it depends on how a CCBH acquires  mass and how it  interacts with the rest of the universe. This is beyond the possible cosmological-fluid description of CCBHs since it is about the property of a single BH. The DEBH and GBH pictures need further input to fix this issue. Although imposing momentum conservation can sound like the simplest choice, this, together with the mass increase, leads to preferred-frame effects. Locally, the mass increase violates the particle 4-momentum conservation [$p^\mu p_\mu = -m^2(t)$]; and if 3-momentum (or spatial angular-momentum) is imposed to be conserved while $p^\mu$ is not conserved, there is a break of boost invariance, since only $p_0$ will not be conserved. This is not a problem in itself: since CCBHs are interacting with the universe, their preferred frame  could be the CMB frame, for instance, but this is an additional ingredient to be specified \cite[see also,][for a related discussion on prefered frames]{Avelino:2023rac}. Moreover, the case with momentum conservation is the one with the strongest phenomenological bounds (see G23). Henceforth we only consider the case without angular momentum conservation. If the latter is imposed, constraints can only become stronger for any given~$k$.

G23 finds that, for a BBH system without eccentricity, the evolution of the orbit radius $r$ about the center of mass follows the same corresponding equation as for standard BHs, apart from a correcting factor (with $c=1$),
\begin{equation}
    \frac{dr}{dt} = - \frac{64}{5} \frac{G^3\mu M^2}{r^3} \left(\frac{a}{a_i}\right)^{3k} \, ,
\end{equation}
 where $M = m_1 + m_2$ and $\mu = m_1 m_2/(m_1 + m_2)$. Terms that depend on $a$  derivatives are negligible. By solving the differential equation above for $r(t)$ and using $r(t_m)=0$ and $t_m = t_{\rm d} + t_i$,  where $t_m$ is the merger time and $t_i$ the initial or formation time, one finds
 \begin{equation}
     \int^{t_{\rm d} + t_i}_{t_i} \left(\frac{a(t)}{a_i}\right)^{3k} dt = \tilde t_{\rm d} \, . \label{GhodlaRelationzi}
 \end{equation}
This  is the relation between the delay times of CCBH and the delay times of standard BHs (see G23). 
This relation does not impose momentum conservation, otherwise the exponent $3 k$ should be replaced by $15 k$.

Equation~\eqref{GhodlaRelationzi} provides the value of $t_{\rm d}$ for given $t_{i}$ (assuming $a(t)$, $k$ and $\tilde t_{\rm d}$ are given). Actually, we need  $t_{\rm d}$ for given $z_m$, because the starting point of our analysis is the observational data, not the initial configuration. Hence we simply write, 
\begin{equation}
     \int^{t_m}_{t_m - t_{\rm d}} \left(\frac{a(t)}{a_i}\right)^{3k} dt = \tilde t_{\rm d} \, . \label{GhodlaRelationzm}
 \end{equation}
Using our reference values for $t_{\rm d}$, eq.~\eqref{referenceValues},  it is possible to plot the relation above for different values of the merging redshift $z_m$. This is displayed in the left panel of Fig.~\ref{fig:ghodla}. As can be seen in this panel, in the small $t_{\rm d}$ region, a given $\tilde t_{\rm d}$ interval is mapped into a $t_{\rm d}$  interval  with approximately the same length. However, for large $t_{\rm d}$ values, a wide interval in the $\tilde  t_{\rm d}$ axis corresponds to a much narrower interval in the $t_{\rm d}$ one. This means that the probability of large $t_{\rm d}$ values is increased with respect to the $\tilde t_{\rm d}$ distribution. The $t_{\rm d}$ distribution for different $z_m$ values is shown in detail in the right plot of the same figure.

\begin{figure}
	\begin{tikzpicture}
  		\node (img1)  {\includegraphics[width=0.44\textwidth]{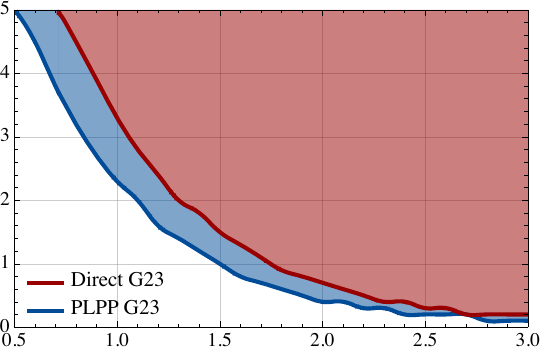}};
		\node[below=of img1, node distance=0cm, yshift=1.1cm, xshift=0.3cm, font=\color{black}] {$k$};
		\node[left=of img1, node distance=0cm, rotate=90, yshift=-0.7cm, xshift=1.6cm] {Minimum BH mass ($M_\odot$)};
	\end{tikzpicture}
	\caption{ 
    {Similar to Figs.~\ref{fig:plotkXmDirect} and \ref{fig:plotkXmPLPP}, but with $t_{\rm d}$ adjusted for CCBH and only considering current data.}
    Excluded regions at 2$\sigma$ using either the direct (red) or the PLPP (blue) method. 
    Both curves consider the uncertainties, either in the observational data (direct) or in the PLPP parameters (PLPP).}
	\label{fig:plotkXmG23} 
\end{figure}

Fig.~\ref{fig:plotkXmG23} combines the direct and PLPP $2\sigma$ bounds on the $k$ and the minimum BH mass constraints, it should be compared with Figs.~\ref{fig:plotkXmDirect} and~\ref{fig:plotkXmPLPP}. These bounds consider the uncertainties in the observational data and on the PLPP parameters. The constraints are much stronger using the G23 relation. We stress here the same particular cases that were considered before: $i$) assuming $m_{\rm th} = 2 M_\odot$ for the minimum BH mass: $k < 1.3$  (direct), $k < 1.1$ (PLPP). $ii$) for $k = 3$, $m_{\rm th} < 0.2 M_\odot$ (direct),  $m_{\rm th} < 0.1 M_\odot$ (PLPP). For $k=1$, $m_{\rm th} < 3.3 M_\odot$ (direct), $m_{\rm th} < 2.3 M_\odot$ (PLPP). For $k = 0.5$, there are no constraints.

\section{Conclusions} \label{sec:conclusions}

According to a recent proposal \citep{Croker:2019mup,Farrah:2023opk},  BHs grow in mass  due to a ``cosmological coupling'' and might be responsible for the cosmic acceleration. In such a scenario, dubbed cosmologically coupled BHs, or CCBH, BHs are not  of the Kerr type, and are supposed to match asymptotically the cosmological  background. This bold idea seems supported by recent analyses of the growth of supermassive BHs in quiescent elliptical galaxies \citep{2023ApJ...943..133F}.

In this paper we tested this hypothesis by considering the binary BHs with stellar progenitors observed with gravitational waves by the LIGO-Virgo-KAGRA (LVK) detectors. If these BHs are cosmologically coupled, they should have undergone a fast mass growth correlated with the cosmological scale factor from the moment of their formation until merging, and should have formed therefore with a mass smaller than the merger one. Since according to the current understanding  there is a mass limit below which stellar BHs cannot form,  the presently observed masses could be in conflict with this mass threshold. To compute this conflict, one needs to know the time between the BBH formation and its merger, which is the delay time ($t_{\rm d}$). The standard $t_{\rm d}$ distribution, without cosmological coupling, is the log-uniform one \citep{KAGRA:2021kbb, KAGRA:2021duu}. It is not established what the CCBH impact on the $t_{\rm d}$ distribution would be, since it depends on unknown microphysical details (it may preserve the log-uniform or not). Our main analysis here is based on either the log-uniform distribution or the CCBH-corrected approach of \cite{Ghodla:2023iaz}. The latter increases our constraints since it favours larger delay times, thus increasing the gap between the formation and the merged mass. This seems to be a  general feature of CCBH-corrections to the standard $t_{\rm d}$ distribution, as discussed in Appendix \ref{app:CCBHtdGeneral}. At last, we also study the consequences of a general power-law $t_{\rm d}$ distribution (Appendix \ref{app:changing_td}). 

We developed two methods of analysis, one is directly based on the observed events only (direct method), and the other is based on the merged BBH population distribution taking into account observational bias, which uses the power-law-plus-peak distribution (PLPP method). The PLPP distribution is the standard model for the merged population of BBHs. Once the uncertainties of each method are taken into account, the results are not far from each other. The PLPP approach yields the strongest constraints.

The main result is that the combination of a minimum BH mass with the GWs data from binary compact objects can put strong bounds on the CCBH approach with current data. For $k=3$ and minimum BH mass ($m_{\rm th}$) of 2 $M_\odot$, a tension of $3 \sigma$ or more is found with the current data.  Moreover, we show that such bounds can {quickly be confirmed} with new data from LVK and that CCBH corrections to the delay-time distribution commonly enhance the constraints. Specifically, with the current $m_1$ data, for  $m_{\rm th} = 2 M_\odot$: $k< 2.5 (1.3)$, for the direct method, and $k < 2.1 (1.1)$  for the PLPP method, at 2$\sigma$ level. The values in parenthesis use the \citet{Ghodla:2023iaz} delay-time correction. Considering the uncertainties on the nature of CCBHs, we also find the required $m_{\rm th}$  value to eliminate the tensions for $k=3$, and find $m_{\rm th} < 0.5 M_\odot$. This is a remarkably low value for a BH-like object of stellar origin, as discussed in Appendix \ref{app:minimumMass}.
Finally, for the CCBH variation studied by \citet{Croker:2021duf}, whose BH masses increase more slowly ($k=0.5$), we found no relevant tension with minimum BH masses. This conclusion is also valid for the case $k=1$, which was recently studied by \citet{Cadoni:2023lum, Cadoni:2023lqe}. 

At last, we stress that If the LVK BHs are of primordial origin, then a completely different analysis would be needed \citep[see also][]{Ghodla:2023iaz}.

\section*{Acknowledgements}
We thank Kevin Croker and Valerio Faraoni for very useful comments and feedback. We also thank Riccardo Sturani for several discussions about this project. LA acknowledges support from DFG project  456622116. 
DCR thanks Heidelberg University for hospitality and support, he also acknowledges support from \textit{Conselho Nacional de Desenvolvimento Científico e Tecnológico} (CNPq-Brazil) and \textit{Fundação de Amparo à Pesquisa e Inovação do Espírito Santo} (FAPES-Brazil) (TO 1020/2022, 976/2022, 1081/2022).
MQ is supported by the Brazilian research agencies FAPERJ, CNPq (Conselho Nacional de Desenvolvimento Científico e Tecnológico) and CAPES. This study was financed in part by the Coordenação de Aperfeiçoamento de Pessoal de Nível Superior - Brasil (CAPES) - Finance Code 001. We acknowledge support from the CAPES-DAAD bilateral project  ``Data Analysis and Model Testing in the Era of Precision Cosmology''.

 \section*{Data Availability}
 
 The data underlying this article are available in the article, in its online supplementary table and in the codes \url{https://github.com/itpamendola/CCBH-direct} and \url{https://github.com/davi-rodrigues/CCBH-Numerics}.


\appendix
\section{CCBH and minimum mass} \label{app:minimumMass}

Here we discuss the adopted $2 M_\odot$ as the main value for the minimum BH mass in the CCBH context. We also add considerations about possible future confirmations of low mass BHs. 

Neutron star (NS) stability studies based on the Tolman-Oppenheimer-Volkoff (TOV) equation predict that nonrotating NS could have masses at least as high as $2.2 M_\odot$ \citep{Ye:2022qoe,Legred:2021hdx}, which sets a lower bound for {Kerr} BHs forming through stellar collapse {(this bound need not  be satisfied for general horizonless compact objects)}. Incidentally, rotating NS have been measured with masses as high as $2.7 M_\odot$~\citep{Romani:2012rh}, and studies of binary systems containing NS find an empirical upper limit as high as $2.6 M_\odot$~\citep{Rocha:2021zos}. This constraint of $2.2 M_\odot$ was used in the main results of \citet{Rodriguez:2023gaa} and \citet{Andrae:2023wge}.

Since GBHs simply grow proportionally to $a^k$,  they may or may not have a horizon. On the other hand, DEBHs  \citep[which are a type of GEODE,][]{Croker:2019mup} need be a source of DE, thus such objects are not expected to have a horizon. They could be named non-singular BHs or exotic compact objects (ECOs) \citep{Mazur:2015kia, Cardoso:2019rvt}. The limit of $2.2 M_\odot$ need not to apply to them \citep[e.g.,][]{Croker:2019mup, Farrah:2023opk, Rodriguez:2023gaa}.

Although non-singular BHs may form from the stellar evolution \citep[e.g.,][]{Mazur:2015kia, Mazur:2001fv} and they may avoid the above horizon-based mass bound, they need to be compatible with the observed NSs with masses as high as $\sim 2.5 M_\odot$, as above mentioned. BH-like objects with masses about one solar mass are not {commonly} expected to be of stellar origin. {For instance,  \citet{LIGOScientific:2022hai} study the possible detection of such objects with LVK, but they are considered to be either of primordial cosmological origin (PBHs) or arising from the collapse of dark matter (which is possible for some particular dark matter models)}. 
In the literature on BHs and NSs, there is a large discussion on the mass gap that separates the maximum non-rotating NSs masses from the minimum detected BH ones. At least currently, there is no support for the generation of stellar BHs with about one solar mass or below. Considering the current scenario, $2 M_\odot$ is a conservative minimum BH mass (i.e., lower than other standard limits), even in the DEBH scenario, considering the lack of a microphysical theory to generate such low mass BHs from stellar evolution.\footnote{{\citet{Gao:2023keg} consider $k>0$ to  explain the mass gap.}} Nonetheless, we also do several specific tests with other minimum mass values.

\begin{figure}
    \begin{tikzpicture}
        \node (img1)  {\includegraphics[width=0.45\textwidth]{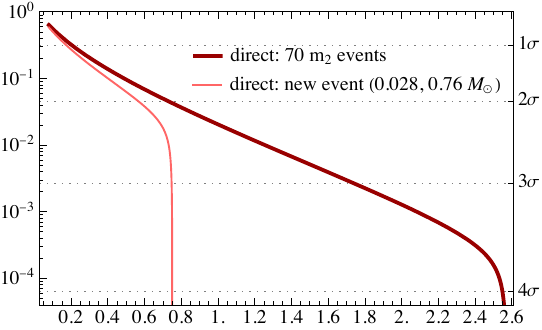}};
        \node[below=of img1, node distance=0cm, yshift=1.1cm, xshift=0.3cm, font=\color{black}] {Minimum mass threshold $m_{\rm th}$ ($M_\odot$)};
        \node[left=of img1, node distance=0cm, rotate=90, yshift=-0.9cm, xshift=0.9cm] {Probability};
    \end{tikzpicture}
    \caption{
    {Same as Fig.~\ref{fig:min-mass-debh} but using $m_2$ data and only for $k=3$.} We use two different data sets: one with 70 $m_2$ events from Table \ref{tab:zXm}, including the event with $m_2 = 2.6 \, M_\odot$ and a second data set that adds the \citet{Morras:2023jvb} object as a stellar BH with $m_2 = 0.76 \, M_\odot$ and $z_m = 0.028$.}
    \label{fig:plotNewLowMassEvent} 
\end{figure}

We consider now the consequences of a low mass BH detection. If observed BHs are assumed to be a primordial one, then our analysis would not apply, since the $t_{\rm d}$ distribution should be considerably different. We only consider BHs of stellar origin. \citet{Morras:2023jvb} present moderate significance evidence for a subsolar compact object detection of mass $m_2 = 0.76^{+ 0.50}_{-0.14} M_\odot$ at $z_m = 0.028$ (it is the low mass companion of the binary system). This is not a strong evidence, since there are still 16\% of chance of its mass being above 1 $M_\odot$, and hence that it could be a NS. Nonetheless, the consequences of a confirmation would be remarkable. Possibly this can be achieved  with either the next LVK run or    the next generation of GW detectors, as the Einstein Telescope \cite[e.g.,][]{Wolfe:2023yuu}.  
If such a small mass compact object  exists and is indeed a  BH-like object of stellar origin, it is possible to infer an upper bound on the minimum BH mass value using our methods. Here we only consider the direct method, since we are only adding a new event and we are not assuming that we know the population of merging low mass BHs.

For comparison, we consider two sets of $m_2$ data. The first set is composed by all the 72 $m_2$ events listed in Table \ref{tab:zXm}, apart from the two least massive ones, which are assumed to be NSs by LVK. For this set, the least massive BH has a mass of $2.6 M_\odot$. The second set is composed by the previous data plus the \citet{Morras:2023jvb} object, with mass $0.76 M_\odot$.

For the first data set, and for $k=3$, the minimum BH mass threshold should satisfy $m_{\rm th} < 0.73 \,M_\odot$ at 2$\sigma$ level. Once the \citet{Morras:2023jvb} object is considered, one finds $m_{\rm th} < 0.57 \,M_\odot$. Figure \ref{fig:plotNewLowMassEvent} shows further details. {In this figure, as the threshold becomes closer to the detected BH mass, the probability quickly goes to zero.}

The above application shows that, as expected, the less massive the detected BH, the smaller should be the minimal CCBH mass to avoid tension with the observational data. To compute a rejection level for given $k$, one needs a reasonable theoretical estimate for a minimum stellar BH mass. For the particular case considered above, if such estimate is below $0.57 M_\odot$, there would be no tension with the current data for $k=3$ (neglecting detection bias).

The delay time distribution has a central role for the constraints we find. The next two appendices are devoted to the $t_{\rm d}$ distribution.

\section{CCBH-modified delay-time distributions compared with the log-uniform one: general approach} \label{app:CCBHtdGeneral}

Here we consider the impact of CCBH on the delay-time distribution under general considerations that  only depend on certain basic qualitative expectations. We show that the CCBH correction to the delay time distribution is not expected to decrease the chances of finding larger $t_{\rm d}$ values with respect to the log-uniform distribution.  It is shown that the particular correction that preserves the log-uniform distribution is a power-law correction, while exponential corrections and polynomials will increase the likelihood of high $t_{\rm d}$ values. In  Sec.~\ref{sec:ghodla} the particular realization studied by \cite{Ghodla:2023iaz} is considered.

Considering CCBHs as true for some fixed $k$ value ($0 < k \leq 3$), and for a given  BBH configuration with given initial masses and initial orbit, one can in principle compute both the expected physical delay time $t_{\rm d}$ within this CCBH framework and an auxiliary delay time $\tilde{t}_{\rm d}$ that ignores the cosmological coupling effects ($k=0$). These delay times are expected to satisfy ${t}_{\rm d} < \tilde{t}_{\rm d}$ since the cosmological coupling increases the BHs masses (thus it reduces the time needed from BBH formation up to the merger). More precisely, we will consider the following hypotheses that naturally emerge from this setting: $i$) there exists a function $T$ that maps physical delay times into auxiliary ones, and this function has an inverse: $ \tilde t_{\rm d} = T(t_{\rm d})$ and $t_{\rm d} = T^{-1}(\tilde t_{\rm d})$. Since the physical delay time is always smaller than the corresponding auxiliary one, $t_{\rm d} < T(t_{\rm d})$. $ii$) Since the CCBH effects are expected to become larger as the delay-time increases, $T(t)$ needs to increase faster than linearly. $iii$) The minimum and maximum limits of the physical delay-time $t_{\rm d}$ are assumed to be known from physical considerations (e.g., considering the maximum redshift at which BBHs can be formed, simulations on the minimum physical time that stellar-formed BBHs need to merge, etc...). They are denoted by $t_{\rm min}$ and $t_{\rm max}$, respectively.  $iv$) The auxiliary delay time is expected to describe the physics with $k=0$, thus its distribution is assumed to be log-uniform and given by the PDF
\begin{equation} \label{tildetPDF}
  \tilde \pi(\tilde t_{\rm d}) = \frac{1}{\ln \left( {\tilde t_{\rm max}}/{\tilde t_{\rm min}}\right) } \frac{1}{\tilde t_{\rm d}} \, ,
\end{equation}
where $\tilde t_{\rm max} \equiv T(t_{\rm max})$ and $\tilde t_{\rm min}\equiv T(t_{\rm min})$ are the maximum and minimum values of the log-uniform distribution. Since $\tilde t_{\rm d}$ is simply an auxiliary quantity, computed assuming that $k=0$, $\tilde t_{\rm max}$ can be larger than the age of the universe. 

The four considerations above are more general than the specific assumptions of \citet{Ghodla:2023iaz} and they  are sufficiently precise to yield that $\log t_{\rm d}$ is not flatly distributed, apart from a specific case, and in general the $\log t_{\rm d}$ is crescent. Indeed, since there is a continuous one-to-one map between $t_{\rm d}$ and $\tilde t_{\rm d}$, the probability of finding a physical $t_{\rm d}$ between the values $t_{\rm min}$ and $t$ can be found from $\tilde \pi$,
\begin{equation}
    P(t) = \int_{t_{\rm min}}^{t} \pi(t_{\rm d}) \, {\rm d} t_{\rm d} =\int_{\tilde t_{\rm min}}^{\tilde t} \tilde \pi(\tilde t_{\rm d}) \, {\rm d} \tilde t_{\rm d}, 
\end{equation}
where $\pi(t_{\rm d})$ is the \textit{a priori} unknown PDF of $t_{\rm d}$ and $\tilde t = T(t)$. Therefore, using eq.~\eqref{tildetPDF}, followed by a redefinition of units,
\begin{equation} \label{probabilityP}
    P(t) = \frac{1}{\ln \left( {\tilde t_{\rm max}}/{\tilde t_{\rm min}}\right) } 
    \ln \left( \frac{T(t)}{\tilde t_{\rm min}}\right)  = 
    \frac{\ln T(t)}{\ln  {\tilde t_{\rm max}} } \, .
\end{equation}
About the units redefinition, one can measure $t_{\rm max}$ and $t$ in units of $t_{\rm min}$, which is in practice equivalent to $t_{\rm min} = 1$. Moreover, since eq.~\eqref{probabilityP} is invariant under $T(t) \to k T(t)$ and $\tilde t_{\rm min} = T(1)$, one can set $\tilde t_{\rm min} =1$ in this equation. These units transformations are not necessary, but may be helpful to understand the essence of the argument.

Now we compare the above probability to a log-uniform probability defined within $1=t_{\rm min} < t < _{\rm max}$ and denoted by $\bar P(t)$,
\begin{equation} \label{PbarP}
    \frac{P(t)}{\bar P(t)} = 
    \frac{\ln  { t_{\rm max}}}{\ln  {\tilde t_{\rm max}} } \frac{\ln T(t)}{\ln t }\, .
\end{equation}
If $\ln T(t)$ increases faster than $\ln t$ for all $t$, then ${P(t)}/{\bar P(t)}$ monotonically increases, implying that $\max [{P(t)}/{\bar P(t)}] = {P(t_{\rm max})}/{\bar P(t_{\rm max})} =  1$. That is, $P(t) < \bar P(t)$ for any $t < t_{\rm max}$. This implies that the $t_{\rm d}$ distribution will not be log-uniform and that  it will favour larger $t_{\rm d}$ values with respect to the log-uniform case. 

A special $T(t)$ function that is in agreement with $\partial_t [T(t)/t] > 0$ (as implied by the item $ii$ above) but $\partial_t [\ln T(t)/\ln t] = 0$ is $T(t) = t^n$, with $n > 1$ and $ 1 < t < t_{\rm max}$. For this case, although $T(t)$ increases faster than $t$, $\ln T(t)$ increases proportionally to $\ln t$, implying a constant $P/\bar P$ ratio. Indeed, computing eq.~\eqref{PbarP} directly,  $\ln \tilde t_{\rm max} = n \ln t_{\rm max} $  and $\ln  T (t)  = n \ln t $, hence ${P(t)}/{\bar P(t)} = 1$. A CCBH correction that can be approximated by $T(t) = t^n$ will not change the log-uniform distribution. Exponentials and polynomials, for example, when used to build monotonically increasing $T(t)$ functions, will change the delay-time distribution and will increase the odds of large delay times with respect to the log-uniform case.

We  remark that $T(t)$ functions that  satisfy the item $ii$ (i.e., $\partial_t [T(t)/t] > 0$), cannot lead to $\partial_t [\ln T(t)/ \ln t] < 0$. 

Provided that the four ($i$-$iv$) general hypothesis listed above are satisfied, and that a log-uniform distribution is a good approximation for the standard ($k=0$) case, we conclude that the CCBH correction to the standard delay time should either preserve the log-uniform distribution and our results from the Secs.~\ref{sec:direct} and \ref{sec:powerLawPlusPeak}, or strength the constraints we find. 

\bigskip 
\section{Changing the delay time distribution slope} \label{app:changing_td}

The delay time distribution is a crucial assumption, so we discuss here briefly the impact of changing the  $1/t_d$ slope. This case is different from the one studied in Appendix \ref{app:CCBHtdGeneral}, since here we do not consider CCBH corrections to the log-uniform distribution, but a direct change in the $t_{\rm d}$ distribution.

\begin{figure}
    \centering
     \includegraphics[width=0.37\textwidth]{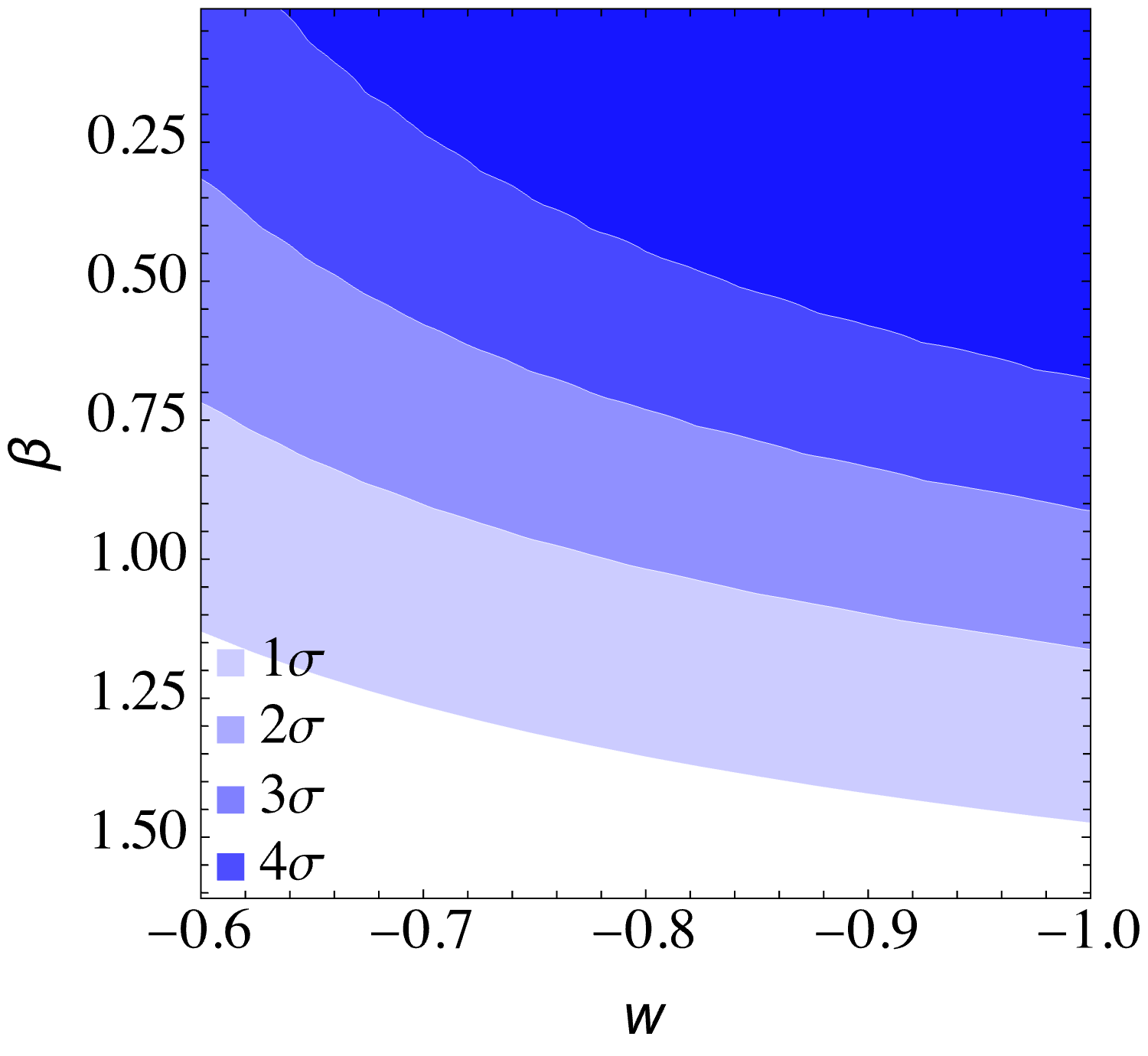}
    \caption{Direct method. Contour plot of $P(m_1>m_{\rm th})$ as a function of $w$ and the slope $\beta$ in the DEBH model. 
    } 	\label{fig:contours-w-beta} 
\end{figure}

The full functional form of the relation delay times versus mass is unknown and it appears difficult to model exactly. Here we limit ourselves to a preliminary investigation.
For this appendix we adopt the more conservative direct method of Sec.~\ref{sec:direct}.  \citet{vanSon:2021zpk} show a tendency for low-mass BHs to be formed via the common envelope channel, and this channel would have a  delay time distribution steeper than $1/t_d$ (i.e., shorter delay times on average), corresponding to $\beta=1.1-1.3$. 
Looking at the results of App.~B of~\citet{vanSon:2021zpk}, we see that such a steeper power-law distribution  approximates the predicted behavior for  masses below $30M_\odot$. For simplicity, we assume that this power-law extends to all  masses: this has anyway very little impact since large masses are not the main drivers of our statistics.  In Fig.~\ref{fig:contours-w-beta} we show the probability contour plot for $w,\beta$. Negative $\beta$ seems totally excluded. For $\beta\ge 1.2$, the probability for $w=-1$ decreases below 2$\sigma$, bringing the DEBH model  into  the non-rejection region. Therefore, as far as current data are concerned, such steeper power-laws might alleviate or solve the tension.

\bsp	
\label{lastpage}
\end{document}